\documentclass{jfm}

\usepackage{graphicx,mathtools,url}
\usepackage{newtxtext}
\usepackage{newtxmath}
\usepackage{natbib}
\usepackage{hyperref}
\hypersetup{
    colorlinks = true,
    urlcolor   = blue,
    citecolor  = black,
}
\newcommand*{\myalign}[2]{\multicolumn{1}{#1}{#2}}
\newcommand{\ut}{\ensuremath{U_{\tau}}}
\newcommand{\utdef}{\ensuremath{U_{\tau}\coloneqq\sqrt{\tau_{w}/\rho}}}
\newcommand{\uinf}{\ensuremath{U_{\infty}}}
\newcommand{\ub}{\ensuremath{U_{b}}}
\newcommand{\utl}{\ensuremath{\widetilde{U}_{\tau}}}
\newcommand{\utlexp}{\ensuremath{U_{\tau}\nu/M}}
\newcommand{\uinftl}{\ensuremath{\widetilde{U}_{\infty}}}
\newcommand{\uinftlexp}{\ensuremath{U_{\infty}\nu/M}}
\newcommand{\ltl}{\ensuremath{\widetilde{L}}}
\newcommand{\ltlexp}{\ensuremath{LM/\nu^{2}}}
\newcommand{\yplus}{\ensuremath{y_{+}}}
\newcommand{\tplus}{\ensuremath{t_{+}}}
\newcommand{\tplusdef}{\ensuremath{t_{+}=t\ut^{2}/\nu}}
\newcommand{\Retau}{\ensuremath{\Rey_{\tau}}}
\newcommand{\ReL}{\ensuremath{\Rey_{L}}}
\newcommand{\Retheta}{\ensuremath{\Rey_{\theta}}}
\newcommand{\Redelta}{\ensuremath{\Rey_{\delta}}}

\newcommand{\cf}{\ensuremath{C_{f}}}

\newcommand{\limit}{\ensuremath{\Rey\rightarrow\infty}}
\newcommand{\limittau}{\ensuremath{\Rey_{\tau}\rightarrow\infty}}
\newcommand{\sqtcfbytwo}{\ensuremath{\sqrt{C_{f}/2}}}
\newcommand{\sqtcfbytwodef}{\ensuremath{\sqrt{C_{f}/2}\coloneqq\ut/\uinf}}
\newcommand{\GbyGref}{\ensuremath{G/G_{\textrm{\emph{ref}}}}}
\newcommand{\Gref}{\ensuremath{G_{\textrm{\emph{ref}}}}}

\newcommand{\RomanNumeralCaps}[1]
\linenumbers

\title{Scaling of mean skin friction in turbulent boundary layers, and fully-developed pipe and channel flows}

\author{Shivsai~Ajit~Dixit
  \corresp{\email{sadixit@tropmet.res.in}},
  Abhishek~Gupta, Harish~Choudhary \and Thara~Prabhakaran}

\affiliation{Indian Institute of Tropical Meteorology, Pune (Ministry of Earth Sciences, New Delhi), India-411008
}

\begin{document}
\maketitle

\begin{abstract}

An asymptotic $-1/2$ power-law scaling and a semi-empirical finite-$\Rey$ model were recently presented by \cite{dixit2020PoF} for skin friction in zero-pressure-gradient (ZPG) turbulent boundary layers (TBLs). In this work, a new derivation is presented which shows that these relations (i) fundamentally represent a dynamically-consistent scaling of skin friction for nominally two-dimensional ZPG TBLs and fully-developed pipes and channels, and (ii) apply individually to each of these flows. The new theoretical arguments are based on transfer of kinetic energy from mean flow to large eddies of turbulence and depend neither on flow geometry nor outer boundary condition, both of which distinguish one type of flow from the other. Using skin friction data from the literature, it is demonstrated that the finite-$\Rey$ model describes, as predicted by the theory, data from individual flows remarkably well; these data cover the complete range of laboratory/simulation Reynolds numbers to date. It is, however, observed that performance of the model degrades while attempting to describe data from all flows in a universal fashion. Differences in outer boundary condition and large-scale structures amongst different types of flows appear to be responsible for this degradation. An empirical correction based on Clauser's shape factor, is proposed to absorb the outer boundary condition effects into the scaling of skin friction. This correction leads to a new universal scaling and a robust, semi-empirical, universal finite-$\Rey$ model for skin friction in ZPG TBLs, pipes and channels. Remarkable collapse of data from all flows in the new scaling underscores the importance of a dynamically-consistent approach towards revealing universality of skin friction in wall turbulence.
\end{abstract}

\begin{keywords}
%Authors should not enter keywords on the manuscript, as these must be chosen by the author during the online submission process and will then be added during the typesetting process (see \href{https://www.cambridge.org/core/journals/journal-of-fluid-mechanics/information/list-of-keywords}{Keyword PDF} for the full list).  Other classifications will be added at the same time.
\end{keywords}

{\bf MSC Codes }  {\it(Optional)} Please enter your MSC Codes here

\section{Introduction}\label{sec:intro}

Ever since the historical pipe flow resistance measurements and formula by Weisbach in the 1840s and Darcy in the 1850s \citep{brown2002}, and the explanations of their results provided by the celebrated work of Osborne Reynolds in the early 1880s \citep{reynolds1883}, the problem of resistance to turbulent flow past solid surfaces has continued to attract research till date. Internal flows through pipes (and to a lesser extent channels) have widespread engineering and industrial applications such as carrying steam from boilers to turbines in thermal and nuclear power plants, and transporting oil and natural gas through transcontinental pipe lines etc. External flows such as boundary layers are of great importance as well, with the applications ranging all the way from the flow over the wings of micro air vehicles, steam or gas flows over the blades of turbines, and flow of air over the wings and fuselage of an aircraft, to the flow of water around the ship hull or past the skin of submarines.

While the interest in turbulent drag of internal flows (pipes and channels) has a long history, the corresponding development on the front of external flows is relatively recent. Foundations of modern fluid dynamics were laid by \cite{prandtl1904} with his revelations on the existence of a boundary layer adjacent to the solid surface in the flow \citep{tani1977}. Subsequent to this breakthrough, the development of fluid dynamics of external flows witnessed rapid growth. The work of Blasius in 1908 \citep[see][for the English translation]{blasius1950} showed that for laminar boundary layer on a flat plate, the dimensionless drag is a function of Reynolds number  - a dimensionless number that had its origins in the pipe flow studies of Osborne Reynolds \citep{reynolds1883}. It was soon realized that the flow of fluid within the boundary layer too undergoes transition from a laminar state to a turbulent state in a manner much similar to that observed in the pipe flow experiments by Reynolds. Very naturally, therefore, it was tempting to connect the dimensionless drag in turbulent boundary layers (TBLs) to Reynolds number consistent with the laminar boundary layer and turbulent pipe flow results. For zero-pressure-gradient (ZPG) TBLs, this connection was established using the universal log law for mean velocity distribution in viscous (inner) and defect (outer) coordinates \citep{coles1955,coles1956,fernholz1996a}; the Clauser chart method \citep{clauser1954}, which uses only the mean velocity log law in inner coordinates, is an offshoot of this approach \citep{dixit2009}. Significant advances were made on the front of experimental techniques for accurate measurement of skin friction in TBLs; the oil film interferometry technique \citep{tanner1976,chauhan2010} is perhaps the most accurate direct method available to date \citep{fernholz1996,chauhan2010} for the measurement of mean skin friction. Even so, little progress was made on the front of scaling of skin friction in TBLs until very recently, a novel scaling was presented by \cite{dixit2020PoF} for ZPG TBLs; the scaling was inspired from the work of \cite{gupta2020} in two-dimensional wall jet flows. The asymptotic scaling law was based on the asymptotic form of the momentum integral equation, and used a new velocity scale $M/\nu$ (derived from the boundary layer kinematic momentum rate $M$ in the streamwise-wall-normal plane and the fluid kinematic viscosity $\nu$) in the formulation of dimensionless drag and flow Reynolds number; we refer to this as the $M$-$\nu$ scaling. It was demonstrated that, finite-$\Rey$ corrections could be incorporated into the asymptotic scaling law to yield a finite-$\Rey$ skin friction model for ZPG TBLs that explained the variation of dimensionless drag with Reynolds number (in the $M$-$\nu$ scaling) quite well over the complete range of Reynolds numbers accessed by experiments/simulations to date.

Over the past three decades or so, exclusive facilities were designed to probe into unprecedented high-$\Rey$ regimes in pipes \citep{zagarola1998JFM} as well as TBLs \citep{nickels2005,vallikivi2015}. Also, the increase in the computational resources enabled probing increasingly higher Reynolds numbers in numerical simulations \citep{lee2015,pirozzoli2021}. The objective was to better understand the asymptotic behaviour of turbulence in these flows. These studies showed that the log-law based models for the variation of friction factor with Reynolds number in pipe flows require adjustment of coefficient values at higher Reynolds numbers \citep{mckeon2005JFM}. It is customary to use the friction factor $\lambda\propto\ut^{2}/\ub^{2}$ in pipe flows where $\ub$ is the bulk velocity, $\utdef$ is the friction velocity wherein $\tau_{w}$ is the wall shear stress and $\rho$ is the density of fluid. Note that, the friction factor $\lambda$ in pipes is equivalent to the skin friction coefficient $\cf\propto\ut^{2}/\uinf^{2}$ in TBLs ($\uinf$ is the freestream velocity), both being dimensionless measures of the drag force per unit area of the surface. In the recent times, the focus gradually shifted from devising phenomenological drag models or formulas that fit the experimental data to a detailed evaluation of the contributions of various structural components or eddies to the drag in wall-bounded turbulent flows. \cite{fukagata2002} derived an integral equation (the so-called FIK identity) that could be used to assess the contribution of the distribution of turbulent shear stress across the thickness of the flow towards the mean skin friction in internal as well as external flows. \cite{deck2014} assessed the contributions of the large-scale structures to the mean skin friction in TBLs within the framework of FIK identity using data from numerical simulations. They showed that the large-scale motions with wavelengths longer than twice the boundary layer thickness \citep[these include `superstructures' in the log region of ZPG TBLs according to][]{hutchins2007a,hutchins2007b,mathis2009} contribute more than $45\%$ of the mean wall shear stress through the footprinting and amplitude modulation effects in the near wall region. \cite{giovenatti2016} concluded that the contribution to the mean skin friction, of the attached coherent motions in the log region of a channel flow, continually increases with Reynolds number until eventually most of the skin friction is contributed by such motions. \cite{gioia2006PRL} showed that the power-law type $\lambda$-$\Rey$ relationship in pipe flows is closely related to the sizes of the eddies (in the turbulence cascade) that cause substantial momentum transfer between the flow and the wall. They showed that the Blasius' $-1/4$ power law scaling is a consequence of the dissipative (Kolmogorov scale) eddies of the cascade effecting most of the momentum transfer between the wall and the fluid layer right next to it. Building upon this approach, recently \cite{anbarlooei2020} argued that a new power-law scaling regime emerges at high Reynolds numbers in pipe flows where the momentum transfer is effected by eddies having sizes of the order of the height of the mesolayer - region of the flow around the Reynolds shear stress maximum in the near-wall region. Very recently, \cite{dixit2021PoF} presented a new universal model for $\lambda$-$\Rey$ relationship in smooth pipes that combines the attached eddy type contributions (typical of the log law models) with the high-wavenumber contributions (typical of the power law models) that are missed out in the attached eddy framework due to its inviscid character. This new universal model was shown to explain the variation of $\lambda$ over the complete range of pipe flow Reynolds numbers at once, without requiring any regime-wise adjustment of coefficients as required by the earlier log law \citep{mckeon2005JFM} and power law \citep{anbarlooei2020} models. 

In the context of scaling of mean skin friction in wall turbulence, two points are worth consideration. First, the scaling of mean skin friction has largely been considered as the by-product of mean velocity scaling laws. For this reason, skin friction laws in the literature typically have the same functional form as the mean velocity overlap layer \citep{fernholz1996a,george1997,zanoun2009refined,mckeon2005JFM}. However, it is important to note that most mean velocity scaling laws are empirical expectations based on a certain physical understanding as to what the correct choice of scales in a certain part of the flow could be. For example, the defect scaling for the outer layer and viscous scaling for the inner layer are essentially empirical in nature i.e they do not follow from the governing equations. Therefore, the overlap layer mean velocity scaling and the corresponding skin friction law, both inherit this unavoidable empiricism. There appear to be no studies that investigate the scaling of skin friction, in its own right and in a \emph{dynamically consistent} manner, without subscribing to any scaling description for the mean velocity field. Secondly, the studies from the past have focussed separately on ZPG TBLs \citep{afzal2001actamech,fernholz1996a}, pipes and channels \citep{afzalyajnik1973,zanoun2007ETFS,zanoun2009refined,mckeon2005JFM}. There have been no attempts to explore the \emph{universality} of the mean skin friction scaling across different types of wall-bounded turbulent flows. This could perhaps be so because the internal and external wall-bounded turbulent flows have very different boundary conditions and outer-layer structural details. Therefore, a step in this direction addressing the two points mentioned above, would present a significant advance in our understanding of the scaling of drag and behaviour of turbulence in different types of flows.

In this work, we approach this problem of a \emph{dynamically-consistent, universal} scaling of skin friction in the context of three canonical flow types namely the ZPG TBL (external flow) and fully-developed pipe and channel flows (internal flows). Henceforth, we shall omit the qualification `fully-developed' for brevity. The outline of the present paper is as follows. In section~\ref{sec:ZPG}, we briefly review the $M$-$\nu$ scaling of skin friction by \cite{dixit2020PoF} for ZPG TBLs. The asymptotic $-1/2$ power-law scaling and the finite-$\Rey$ model derived therein are stated; these results, as per the analysis by \cite{dixit2020PoF}, are valid only for ZPG TBLs. Section~\ref{sec:generalflows} introduces a set of new theoretical arguments that are more generally applicable to ZPG TBLs as well as pipes and channels. These arguments are based on the transfer of mean flow kinetic energy to turbulence, and show that the $M$-$\nu$ scaling (including the asymptotic $-1/2$ power-law scaling and the finite-$\Rey$ model) in fact applies individually to ZPG TBLs, pipes as well as channels. In this section, we also show that the velocity scale $M/\nu$, proposed earlier by \cite{dixit2020PoF} only for ZPG TBLs, now emerges as the dynamically consistent velocity scale for skin friction in all the three types of flows under the present consideration. It is argued that the velocity scales in traditional use may be convenient but not necessarily dynamically consistent. Section~\ref{sec:analysis} presents preliminary analysis of scaling of skin friction data in the traditional $(\Rey,\cf)$ space. This is followed in section~\ref{sec:furtheranalysis} by a detailed scrutiny of the $M$-$\nu$ scaling for individual flows (channels, pipes and ZPG TBLs) in the $(\ltl,\utl)$ space; this scaling is shown to work very well in line with the expectation of the theory presented in section~\ref{sec:generalflows}. However, it is further shown that the $M$-$\nu$ scaling degrades while attempting a universal description of skin friction for all flows. Section~\ref{sec:newscaling} gives the rationale and details of the new universal scaling for all flows. First, the connection between the outer boundary conditions and the large-scale structures in the outer layer of a flow is discussed including structural contributions to the mean skin friction. It is argued that the effects of outer boundary conditions on the mean skin friction may be factored in using the shape of the mean velocity profile which is different in each type of flow. An empirical correction to account for these differences is proposed. The correction utilizes the ratio $\GbyGref$ ($G$ is the Clauser shape factor and $\Gref=6.8$ is its reference value for ZPG TBLs) and leads to a remarkable universal scaling behaviour in the new shape-factor-corrected $(\ltl',\utl')$ space. We refer to this new universal scaling for all flows as the $M$-$\nu$-$G$ scaling. A three-dimensional interpretation of the present $M$-$\nu$-$G$ scaling in terms of the $(\ltl,\GbyGref,\utl)$ space is also discussed. Conclusions are presented in section~\ref{sec:conclusion}.    
 
%In the quest for a universal skin friction scaling, surely it is not unreasonable to expect that the differences in the outer boundary conditions of different types of flows could play crucial roles. For example, while the ZPG TBL possesses the freestream outer boundary condition, a eloped pipe or channel flow is turbulent throughout its core with no freestream boundary. These differences reflect in the mean velocity profiles of as differences the outer regions of these flows are known to be different.

\section{$M$-$\nu$ scaling of skin friction in ZPG TBLs}\label{sec:ZPG}

\subsection{Asymptotic $-1/2$ power law}\label{subsec:ZPG-asymplaw}

Recently, \cite{dixit2020PoF} have considered the asymptotic form of the integral momentum equation for ZPG TBLs where the flow is two-dimensional in the mean i.e only streamwise ($x$) and wall-normal ($y$) variations are important. They show that, an asymptotic skin friction law, previously unrecognised, may be derived wherein the local (at a particular $x$ location) friction velocity $\ut$ and boundary layer thickness $L$ are made dimensionless using an a dynamically-relevant local velocity scale $M/\nu$ instead of the traditional local freestream velocity scale $\uinf$. Here, $\nu$ is the fluid kinematic viscosity and $M=\int_{0}^{L}U^{2}\textrm{d}y$ is the boundary layer kinematic momentum rate (henceforth, simply momentum rate for brevity) in the $x$-$y$ plane (per unit width in the spanwise i.e. $z$ direction) at a streamwise location $x$. Note that, $U=U(y)$ is the mean velocity profile at the streamwise location $x$. The quantities $\ut$, $M$ and $L$ are functions of the streamwise coordinate $x$ and in what follows, we shall omit explicit mention of $x$ with an understanding that the quantities being considered are localised in the streamwise direction unless specified otherwise. The dimensionless friction velocity and boundary layer thickness in the formulation of \cite{dixit2020PoF} are $\utl\coloneqq\utlexp$ and $\ltl\coloneqq\ltlexp$ respectively ($\coloneqq$ stands for `by definition'). We refer to this non-dimensionalization as the $M$-$\nu$ \emph{scaling} because the ZPG TBL data \emph{collapse} to a universal curve when plotted in the $(\ltl,\utl)$ space \citep[see figure~2\emph{a} of][]{dixit2020PoF}. In the limit $\limit$, the asymptotic skin friction law in the $M$-$\nu$ scaling takes the form of a $-1/2$ power-law relationship
\begin{equation}
\utl\sim\ltl^{-1/2}.\label{eqn:asymplaw1}
\end{equation}
\noindent Note that, the traditional scaling of skin friction uses, as mentioned earlier, the freestream velocity $\uinf$ as the velocity scale. Thus, in the traditional framework, the dimensionless friction velocity is $\ut/\uinf\coloneqq\sqtcfbytwo$ ($\cf$ is the skin friction coefficient) and the dimensionless boundary layer thickness is the Reynolds number $\ReL\coloneqq L\uinf/\nu$ based on $\uinf$ and $L$. Comparing these with the definitions of $\utl$ and $\ltl$ indicates that $\utl$ is akin to $\sqtcfbytwo$, and $\ltl$ is akin to $\ReL$ when $\uinf$ is replaced by the velocity scale $M/\nu$.

\subsection{Finite-$\Rey$ model}\label{subsec:ZPG-finiteRemodel}

As mentioned in section~\ref{subsec:ZPG-asymplaw} before, \cite{dixit2020PoF} have shown that the variables $\utl$ and $\ltl$ lead to remarkable scaling of the ZPG TBL data over the complete range of Reynolds numbers accessed to date using simulations and experiments. Further, it has been shown that, the $-1/2$ power-law (\ref{eqn:asymplaw1}) is valid only in the limit $\limit$ and finite-$\Rey$ corrections are required to obtain the correct functional form that describes the data (in the $M$-$\nu$ scaling) over the complete range of (finite) Reynolds numbers. Towards this, Dixit \etal~have proposed to write the skin friction law more generally as $\utl=A\ltl^{B}$. The coefficient $A$ and exponent $B$ have then been expanded in terms of semi-empirical asymptotic series expansions such that, $A\rightarrow0$ and $B\rightarrow-1/2$ in the limit $\limit$. Retaining the terms in these expansions up to the first order, Dixit \etal~obtain a semi-empirical finite-$\Rey$ model for skin friction in ZPG TBLs 
\begin{equation}
\utl=\frac{A_{1}}{\ln\ltl}\ltl^{\left[-\frac{1}{2}+\frac{B_{1}}{\sqrt{\ln\ltl}}\right]},\label{eqn:finitelaw1}
\end{equation}
\noindent where $A_{1}$ and $B_{1}$ are empirical constants to be obtained by fitting (\ref{eqn:finitelaw1}) to the data. Equation (\ref{eqn:finitelaw1}) implies that $\utl$ is an explicit function of $\ltl$ with the general form
\begin{equation}
\utl=F_{1}(\ltl),\label{eqn:finitelaw1generalform}
\end{equation}
\noindent where the functional form of $F_{1}$ is given by the right side of (\ref{eqn:finitelaw1}). Notice that, $\ltl$ is known from the measured or computed mean velocity distribution across the TBL height, whereas $\utl$ is an unknown to be obtained from the knowledge of $\ltl$. Dixit \etal~have shown that (\ref{eqn:finitelaw1}) describes the variation of $\utl$ over the complete range of $\ltl$ to an excellent accuracy; more details of the derivation of (\ref{eqn:finitelaw1}) can be found in their paper.

\section{$M$-$\nu$ scaling of skin friction in ZPG TBLs, pipes and channels}\label{sec:generalflows}

The chief difficulty with (\ref{eqn:asymplaw1}) is that it applies only to the ZPG TBLs. This is because (\ref{eqn:asymplaw1}) has been formally derived from the asymptotic form of the integral momentum equation for ZPG TBLs \citep{dixit2020PoF}. It would however, be interesting to see if (\ref{eqn:asymplaw1}) applies more generally to other wall-bounded turbulent flows as well. Of particular interest to us in this work are the ZPG TBLs, pipe and channel flows, since these three flows constitute the canonical flow archetypes of wall turbulence and are the most extensively studied in the literature. There are, however, two fundamental differences amongst these three flow types. First, the nature of the outer boundary condition is different; ZPG TBLs being external flows have a freestream whereas pipes and channels are internal flows and do not possess a freestream. Second, the flow geometry is different; ZPG TBLs and channel flows naturally conform to the Cartesian coordinates whereas cylindrical coordinates are apt for pipe flows. However, under the assumption of spanwise homogeneity for ZPG TBLs and channels, and azimuthal homogeneity for pipes, the mean-flow governing equations for all of them become identical and two-dimensional in the streamwise-wall-normal plane \citep{schlichting1968,kundu2008,davidson2015}. Therefore, with these assumptions, differences in the flow geometry pose no hurdles. Thus, it would be of interest to examine if (\ref{eqn:asymplaw1}) applies asymptotically without being influenced by the differences in the outer boundary condition that distinguish one type of flow from the other. This is possible if (\ref{eqn:asymplaw1}) can be derived using a theoretical approach (in the limit $\limit$) that applies to all the wall-bounded turbulent flows of the present interest irrespective of the differences in the outer boundary condition.

The mechanism of the transfer of kinetic energy from mean flow to turbulence by the large eddies of the flow is a universal feature of turbulent wall-bounded shear flows that does not depend on the boundary conditions. Hence, in what follows, we propose a new kinetic energy transfer argument which is based on the following three key facts.
\begin{enumerate}
\item The source term in the governing equation for TKE is the sink term in the governing equation for the streamwise mean-flow kinetic energy (SMFKE) - a well-known mechanism by which shear-flow turbulence extracts energy from the mean flow through the work done of turbulent shear stresses on the mean velocity gradient \citep{tennekes1972,davidson2015}. 
\item For ZPG TBLs, pipes and channels, the average rate of turbulence kinetic energy (TKE) production over the characteristic thickness $L$ of the shear flow may be shown to asymptote to $\ut^{3}/L$ in the limit $\limit$.
\item The eddies that are most efficient in converting the SMFKE to the TKE have their sizes scaling on the thickness $L$ of the shear flow \citep{davidson2015}.
\end{enumerate}

In the following, we elaborate on these aspects and proceed to show that the asymptotic $-1/2$ power-law (\ref{eqn:asymplaw1}) applies to all wall-bounded turbulent flows of the present interest.  

\subsection{Integral SMFKE equation and the `process' of loss of SMFKE to TKE}\label{subsec:generalflows-SMFKEtoTKE}

Consider a wall-bounded turbulent shear flow with characteristic thickness $L$ in the wall-normal direction. Note that, $L$ is the boundary layer thickness, pipe radius and channel-half height for ZPG TBLs, pipe flows and channel flows respectively. Due to the presence of the wall, the problem involves two length scales, namely the viscous length scale $\nu/\ut$ governing the near wall (viscous) dynamics of turbulence and the outer length scale $L$ dictating the sizes of the largest (inertial) eddies of turbulence. The ratio of these two length scales is the friction Reynolds number $\Retau\coloneqq L\ut/\nu$. The streamwise mean momentum equation for a nominally two-dimensional and statistically stationary flow (under the boundary layer approximation for external flows), reads
\begin{equation}
U\frac{\p U}{\p x}+V\frac{\p U}{\p y}=-\frac{1}{\rho}\frac{\textrm{d}p}{\textrm{d}x}+\nu\frac{\p^{2}U}{\p y^{2}}+\frac{\p\left<-u'v'\right>}{\p y}.\label{eqn:xmom}
\end{equation}
\noindent Here, $U$ and $V$ are the mean velocities in the streamwise ($x$) and wall-normal ($y$) directions respectively. $\textrm{d}p/\textrm{d}x$ is the mean streamwise pressure gradient, $u'$ is streamwise velocity fluctuation, $v'$ is the wall-normal velocity fluctuation, and $\left<-u'v'\right>$ is the Reynolds shear stress (pointed brackets denote time average). Note that, (\ref{eqn:xmom}) applies to ZPG TBLs and channel flows with the assumption of spanwise homogeneous mean flow. For pipe flows, (\ref{eqn:xmom}) holds under the assumption of azimuthally homogeneous mean flow. For pipes and channels, the left side of (\ref{eqn:xmom}) goes to zero due to the fully-developed nature of the flow. For ZPG TBLs, the first term on the right side of (\ref{eqn:xmom}) is zero. However, we shall retain all the terms to preserve generality and discuss the implications of some of them being zero in pipes, channels and ZPG TBLs after the general integral SMFKE has been derived. Multiplying (\ref{eqn:xmom}) throughout by $U$ and rearrangement yields the SMFKE
\begin{equation}
\left[U\frac{\p}{\p x}+V\frac{\p}{\p y}\right]\left(\frac{U^{2}}{2}\right)=-\frac{1}{\rho}U\frac{\textrm{d}p}{\textrm{d}x}+\frac{\p}{\p y}\left[U\left(\nu\frac{\p U}{\p y}+\left<-u'v'\right>\right)\right]-\left(\left<-u'v'\right>\frac{\p U}{\p y}\right)-\nu\left(\frac{\p U}{\p y}\right)^{2}.\label{eqn:SMFKE0}
\end{equation}
The left side of (\ref{eqn:SMFKE0}) is the advection of SMFKE i.e. the rate of increase of SMFKE due to movement along a mean streamline of the flow. The first term on the right side (denoted henceforth by $\textrm{\emph{PG}}$) is the rate of work done by the pressure gradient force. The second term (henceforth $T$) is the rate of transport of SMFKE by the viscous and turbulent shear stresses. The third term (henceforth $P$) is the rate of loss of SMFKE due to the work done of the turbulent shear stress against the mean velocity gradient; this term is the gain for the TKE and hence called as the TKE production rate term \citep[or simply production,][]{tennekes1972,davidson2015}. The last term (henceforth $D$) is the rate of direct viscous dissipation of the SMFKE; for a turbulent flow, this term is negligibly small compared to all the other terms \citep{tennekes1972,davidson2015} and may therefore, be neglected. Noting that, $U\p/\p x +V\p/\p y=D/Dt$ i.e. the material derivative operator signifying time rate of change following the fluid along a mean streamline, we divide (\ref{eqn:SMFKE0}) throughout by $\ut^{4}/\nu$ to obtain the SMFKE in viscous scaling
\begin{equation}
\frac{D}{D\tplus}\left(\frac{U_{+}^{2}}{2}\right)\approx \textrm{\emph{PG}}_{+}+T_{+}-P_{+},\label{eqn:SMFKE1}
\end{equation}
\noindent where $\tplusdef$ is the dimensionless time coordinate in viscous or wall units (using $\ut$ and $\nu$ for nondimensionalization) and $U_{+}=U/\ut$ is the streamwise mean velocity in viscous scaling; subscript $+$ denotes the viscous scaling.

Next, we integrate (\ref{eqn:SMFKE1}) in the wall-normal direction over the thickness $L$ of the shear flow i.e. from $\yplus=0$ to $\Retau$; $\yplus=y\ut/\nu$ is the distance from the wall in viscous units. The term $T_{+}$, being a divergence term, integrates to zero in this case i.e. $\int_{0}^{\Retau}T_{+}\textrm{d}\yplus=0$. Therefore, we have
\begin{equation}
\int_{0}^{\Retau}\frac{D}{D\tplus}\left(\frac{U_{+}^{2}}{2}\right)\textrm{d}\yplus\approx\int_{0}^{\Retau}PG_{+}\textrm{d}\yplus-\int_{0}^{\Retau}P_{+}\textrm{d}\yplus.\label{eqn:MFKE2}
\end{equation}   

With this, we are now in a position to appreciate the actual \emph{process} of the transfer of SMFKE to TKE, and consider implications of different terms in (\ref{eqn:MFKE2}) being zero for different types of flows. For ZPG TBLs, $PG_{+}=0$ so that, the first integral on the right side of (\ref{eqn:MFKE2}) vanishes identically. This implies that the integral material derivative term on the left side of (\ref{eqn:MFKE2}) must scale as the integral production term on the right side, and the overall rate of SMFKE loss is simply balanced by the overall rate of TKE production. For pipe and channel flows, the situation is somewhat more subtle. Due to the fully-developed character of these flows, the integral material derivative term on the left side of (\ref{eqn:MFKE2}) is mathematically zero and the overall rate of pressure gradient work balances the overall rate of TKE production. However, it is crucial to recognize that both these balancing terms are forcing agencies in (\ref{eqn:MFKE2}) and for this reason, the production of TKE cannot come directly from the work of pressure gradient force without the SMFKE acting as an intermediate agent (or a medium) through which the communication and balance between the two forcing terms is established. Therefore, the actual process of balance in pipes and channels, consists of two steps that take place simultaneously in time. First, over an infinitesimal period of time, the pressure gradient force does an infinitesimally small work on the SMFKE and causes an infinitesimally small increase in the SMFKE. In the second step, this increase of SMFKE is immediately and simultaneously lost to the infinitesimally small production of the TKE over the same period of time. Thus, although the left side of (\ref{eqn:MFKE2}) is mathematically zero for (fully-developed) pipes and channels, and hence unimportant as one might think, it is dynamically indispensable due to the pivotal role it plays in the process of balance of energy rates of the forcing terms. Hence, for pipes and channels as well, the integral material derivative term on the left side of (\ref{eqn:MFKE2}) must scale as the integral production term on the right side. Thus, for ZPG TBLs, pipes and channels, one may write 
%For TBL flows with pressure gradients, both SMFKE advection and work of pressure gradient force are non-zero. In such cases, a combination of the ZPG and pipe/channel flow arguments stated above, applies. Apparently, the only exceptions where the present analysis does not apply are the TBL flows near separation and relaminarization. This is because $\ut\rightarrow 0$ and is irrelevant in the former case \citep{dengel1990} while the mean-velocity log law itself does not hold for the latter \citep{narasimha1979,warnack1998,dixit2010}. Therefore, for all other TBL flows with pressure gradients, the integral material derivative term on the left side of (\ref{eqn:MFKE2}) must scale as the integral production term on the right side.
\begin{equation}
\int_{0}^{\Retau}\frac{D}{D\tplus}\left(\frac{U_{+}^{2}}{2}\right)\textrm{d}\yplus\sim\int_{0}^{\Retau}P_{+}\textrm{d}\yplus,\label{eqn:MFKE3}
\end{equation}
\noindent where $\sim$ stands for `scales as'. In order to proceed further, it is required to determine the asymptotic value of the integral on the right side of (\ref{eqn:MFKE3}) as shown next.

\subsection{Asymptotic average TKE production rate over the thickness $L$ of the shear flow}\label{subsec:generalflows-Pavg}

The TKE production term $P\coloneqq\left<-u'v'\right>\p U/\p y$ in the viscous (wall) scaling is $P_{+}=P\nu/\ut^{4}$. The average value of $P_{+}$ across the thickness of the shear flow (in the $x$-$y$ plane) is
\begin{equation}
P_{+\textrm{\emph{avg}}}=\frac{1}{L}\int_{0}^{L}P_{+} \textrm{d}y=\frac{1}{\Retau}\int_{0}^{\Retau}P_{+} \textrm{d}\yplus,\label{eqn:avgprod1}
\end{equation}
\noindent It is well-known that although $P_{+}$ reaches peak value in the buffer layer at $\yplus\approx12$, the contribution of the inertial overlap layer (log region) to the overall production (i.e. the integral in \ref{eqn:avgprod1}) increases with Reynolds number and dominates over the contribution of the buffer-layer region at high Reynolds numbers \citep{smits2011}. Physically, this happens because with increasing Reynolds number, the lower end of the log region moves closer to the wall in the outer scaling $(\eta\coloneqq y/L)$ with the buffer layer region sandwiched to an increasingly thinner fraction of the flow thickness located adjacent to the wall; the outer end of the log region remains located at $\eta\approx0.15$ independent of the flow Reynolds number \citep{marusic2013}. In view of this, one may split the integral in (\ref{eqn:avgprod1}) into contributions from the buffer layer, log layer and wake layer regions, and retain only the log and wake contributions in the limit $\limit$
\begin{equation}
P_{+\textrm{\emph{avg}}}\rightarrow\frac{1}{\Retau}\left[\int_{3\sqrt{\Retau}}^{0.15\Retau}P_{+\textrm{\emph{log}}}\textrm{d}\yplus+\int_{0.15\Retau}^{\Retau}P_{+\textrm{\emph{wake}}}\textrm{d}\yplus\right].\label{eqn:avgprod2}
\end{equation}
\noindent Here, the log region in ZPG TBLs, pipes and channels is taken to begin at a Reynolds-number-dependent wall-normal location $y_{+}=3\sqrt{\Retau}$ \citep{marusic2013,wei2005} beyond the mesolayer and extend up to the Reynolds-number-independent location $\eta=0.15$ or $\yplus=0.15\Retau$. The wake region occupies the remaining portion of the flow beyond the log region i.e.~$0.15\Retau\leq\yplus\leq\Retau$. For the wake part, TKE production is governed only by the outer length scale $L$ \citep[as in \emph{free} shear flows, see][]{tennekes1972,townsend1976} so that, $P_{\textrm{\emph{wake}}}\sim \ut^{3}/L$ or $P_{+\textrm{\emph{wake}}}=C_{2}/\Retau$ where $C_{2}$ is a dimensionless constant. For the log region, the production term depends on the distance from the wall \ie~$P_{\textrm{\emph{log}}}\sim\ut^{3}/y$  or $P_{+\textrm{\emph{log}}}=C_{1}/\yplus$ \citep{tennekes1972,townsend1976,davidson2015}, $C_{1}$ being a dimensionless constant. Substituting the expressions for $P_{+\textrm{\emph{wake}}}$ and $P_{+\textrm{\emph{log}}}$ into (\ref{eqn:avgprod2}) and simplifying yields
\begin{equation}
P_{+\textrm{\emph{avg}}}\rightarrow\frac{1}{\Retau}\left[C_{3}+C_{1}\ln{\sqrt{\Retau}}\right],\label{eqn:avgprod3}
\end{equation}
\noindent where $C_{3}=-2.9957C_{1}+0.85C_{2}$. In order to obtain the correct asymptotic limiting form of (\ref{eqn:avgprod3}), one needs to consider the Reynolds-number-dependence of the fractional change $\textrm{d}P_{+\textrm{\emph{avg}}}/P_{+\textrm{\emph{avg}}}$. This may be easily done by differentiating (\ref{eqn:avgprod3}) with respect to $\Retau$ and dividing the result by (\ref{eqn:avgprod3}). With some simplifications, this exercise yields
\begin{equation}
\frac{\textrm{d}P_{+\textrm{\emph{avg}}}}{P_{+\textrm{\emph{avg}}}}\rightarrow\left[-1+\frac{C_{1}}{2\left(C_{3}+C_{1}\ln\sqrt{\Retau}\right)}\right]\frac{\textrm{d}\Retau}{\Retau},\label{eqn:avgprod4}
\end{equation}
\noindent where the square bracket tends to $-1$ in the limit $\limittau$ (or $\limit$). Therefore, (\ref{eqn:avgprod4}) asymptotically becomes 
\begin{equation}
\frac{\textrm{d}P_{+\textrm{\emph{avg}}}}{P_{+\textrm{\emph{avg}}}}\rightarrow-\frac{\textrm{d}\Retau}{\Retau},\label{eqn:avgprod5}
\end{equation}
\noindent which, in turn, shows that
\begin{equation}
P_{+\textrm{\emph{avg}}}\rightarrow\frac{1}{\Retau}\textrm{\hspace{8pt} or \hspace{8pt}}P_{\textrm{\emph{avg}}}\rightarrow\frac{\ut^{3}}{L}.\label{eqn:avgprod6}
\end{equation}
\noindent Thus, the rate of TKE production averaged over the thickness of the shear flow asymptotes to $\ut^{3}/L$ in the limit $\limit$ (or $\limittau$). Notice that, the asymptotic value of $P_{\textrm{\emph{avg}}}$ effectively scales as the value of $P_{\textrm{\emph{wake}}}$ i.e. the functional form (logarithmic) of the mean velocity profile in the inertial overlap layer appears to be irrelevant in the asymptotic sense.

Substituting (\ref{eqn:avgprod1}) and (\ref{eqn:avgprod6}) in (\ref{eqn:MFKE3}) shows that 
\begin{equation}
\int_{0}^{\Retau}\frac{DU_{+}^{2}}{D\tplus}\textrm{d}\yplus\rightarrow\textrm{constant}.\label{eqn:MFKE3extra}
\end{equation} 
\noindent in the limit $\limit$ (or $\limittau$). Furthermore, we note that the limits of integration on the left side of (\ref{eqn:MFKE3extra}) are independent of time (or movement along a streamline) in the limit $\limit$ (or $\limittau$). This is so because, for pipes and channels, the mean flow streamlines are parallel to the wall and coincident with the iso-$\yplus$ lines so that, the limits of integration on the left side of (\ref{eqn:MFKE3}) do not change if one moves along a mean streamline. For ZPG TBLs, this condition is satisfied only at high Reynolds numbers as shown by \cite{dixit2018}. Therefore, in the limit $\limit$ (or $\limittau$), the operators $D/D\tplus$ and $\int$ commute and (\ref{eqn:MFKE3extra}) asymptotically becomes
\begin{equation}
\frac{D}{D\tplus}\int_{0}^{\Retau}U_{+}^{2}\textrm{d}\yplus\rightarrow\textrm{constant}.\label{eqn:MFKE4}
\end{equation} 
\noindent Noting that, the value of the integral in the above equation is simply $M/\nu\ut$, where $M=\int_{0}^{L}U^{2}\textrm{d}y$ is the shear-flow (kinematic) momentum rate in the $x$-$y$ plane (per unit width for ZPG TBLs and channels, and per unit circumference for pipes), one obtains
\begin{equation}
\frac{D}{D\tplus}\left(\frac{M}{\nu\ut}\right)\rightarrow\textrm{constant}.\label{eqn:MFKE5}
\end{equation}

\subsection{Most efficient energy extracting eddies and the asymptotic $-1/2$ power law}\label{subsec:generalflows-mostefficineteddies}

It is well-known that the largest eddies of a wall-bounded turbulent shear flow have sizes of order $L$ and their velocity scale is $\ut$. These eddies are the most efficient towards extracting the SMFKE and transferring it to the low-wavenumber end of the turbulence cascade \citep{tennekes1972,davidson2015}. The lifetime of these large eddies - the so-called large-eddy turnover time - is of the order of $L/\ut$ \citep{tennekes1972}. In viscous units, the large-eddy turnover time is simply $\left(L/\ut\right)\ut^{2}/\nu=\Retau$ - Reynolds number may be interpreted as the ratio of the largest to smallest (viscous) time scales in the flow \citep{tennekes1972}. We now integrate (\ref{eqn:MFKE5}) over one large-eddy turnover time ($\tplus$ from $0$ to $\Retau$) to obtain the SMFKE lost by the mean flow (or the TKE input at the largest flow scales of the cascade), over its complete wall-normal extent, during the lifetime of a typical large eddy. This yields
\begin{equation}
\frac{M}{\nu\ut}\sim\Retau,\label{eqn:MFKE6}
\end{equation} 
\noindent which upon rearrangement immediately leads to the asymptotic $-1/2$ power law for skin friction (\ref{eqn:asymplaw1})
\begin{equation*}
\utl\sim\ltl^{-1/2}.
\end{equation*} 
\noindent Notice that, the outer velocity boundary condition $U(y=L)=\uinf$ has not been used anywhere in the arguments of this derivation. Thus, we have shown that the asymptotic $-1/2$ power law for skin friction (\ref{eqn:asymplaw1}) holds for ZPG TBLs, pipes as well as channels irrespective of the differences in the outer boundary condition.

\subsection{Dynamically consistent velocity scale $M/\nu$}\label{subsec:dynamicallyconsistentscale}

The conceptual implications of the present new derivation of the asymptotic $-1/2$ power law for skin friction (\ref{eqn:asymplaw1}) are quite revealing. First, the derivation shows that the law holds for ZPG TBLs, pipes and channels, and depends upon neither the flow geometry nor the outer boundary condition. Secondly, for the dimensionless representation of the behaviour of skin friction with Reynolds number, $M/\nu$ emerges as a new velocity scale that is consistent with the governing dynamical equations. Thus, the $M$-$\nu$ scaling inbuilt in (\ref{eqn:asymplaw1}) and (\ref{eqn:finitelaw1}) is a dynamically consistent scaling description for all the flows of the present interest. This is very significant because traditionally the skin friction data in external flows (ZPG TBLs) have been `scaled' using the freestream velocity scale $\uinf$ \citep{fernholz1996a,dixit2020PoF} whereas for the internal flows (pipes and channels), the tradition is to use the bulk or the mass-averaged velocity scale $U_{b}$ \citep{mckeon2005JFM,dixit2021PoF}. These velocity scales have been in wide use perhaps because of their role either as the outer `velocity' boundary condition ($\uinf$ in TBLs) or as a practical convenience ($U_{b}$ is proportional to flow rate through the pipe). Given the velocity scale, skin friction coefficient (or friction factor) and Reynolds number, in their traditional from, are simply a consequence of the dimensional analysis following Buckingham's Pi theorem. It is important to realize that these traditional velocity scales are more akin to the outer boundary condition than the governing dynamics. On the other hand, our theory in the preceding sections has, for the first time and to the best of our knowledge, provided a single, dynamically-consistent velocity scale $M/\nu$ for skin friction in ZPG TBLs, pipes and channels. This provides a basis for exploring universality of skin friction scaling amongst these flows.  

\subsection{Finite-$\Rey$ model for skin friction}\label{subsec:newtheoryfiniteRemodel}

As seen above, the asymptotic $-1/2$ power law for skin friction (\ref{eqn:asymplaw1}) can be derived from a set of general arguments based on the fundamental process of conversion of the SMFKE to TKE in wall-bounded turbulent flows. Since, this process is independent of the outer boundary condition of the flow, (\ref{eqn:asymplaw1}) is now seen to be applicable not only to ZPG TBLs but also to pipes and channels. Therefore, the finite-$\Rey$ model (\ref{eqn:finitelaw1}) 
\begin{equation*}
\utl=\frac{A_{1}}{\ln\ltl}\ltl^{\left[-\frac{1}{2}+\frac{B_{1}}{\sqrt{\ln\ltl}}\right]},
\end{equation*}
\noindent is also applicable to each of these flows individually. Although the functional forms of (\ref{eqn:asymplaw1}) and (\ref{eqn:finitelaw1}) remain intact in the case of ZPG TBLs, pipes and channels, the differences in the outer boundary condition could still manifest through the values of model coefficients varying from one flow to the other. %An immediate implication of these theoretical results is that there could be an underlying universal scaling of skin friction in the space formed by the variables $\ltl$ and $\utl$, provided the differences in the outer boundary condition amongst different flows are correctly accounted for. In what follows, we examine these in some detail.

\section{Preliminary analysis of the data: the traditional $(\Rey,\cf)$ space}\label{sec:analysis}

We now assess the skin friction data from the experimental and direct numerical simulation (DNS) studies of ZPG TBLs, pipes and channels available in the literature. These data have been chosen to cover the complete range of Reynolds numbers accessed to date in laboratory and simulation studies and are listed in Appendix~\ref{appA}. First, we shall examine the scaling behaviour in the traditional space of skin friction coefficient ($\cf$) and Reynolds number ($\Rey$).

\subsection{Lack of scaling in the $(\Rey,\cf)$ space}\label{subsec:analysis-traditional}

\begin{figure}
  \centerline{\includegraphics[scale=0.475]{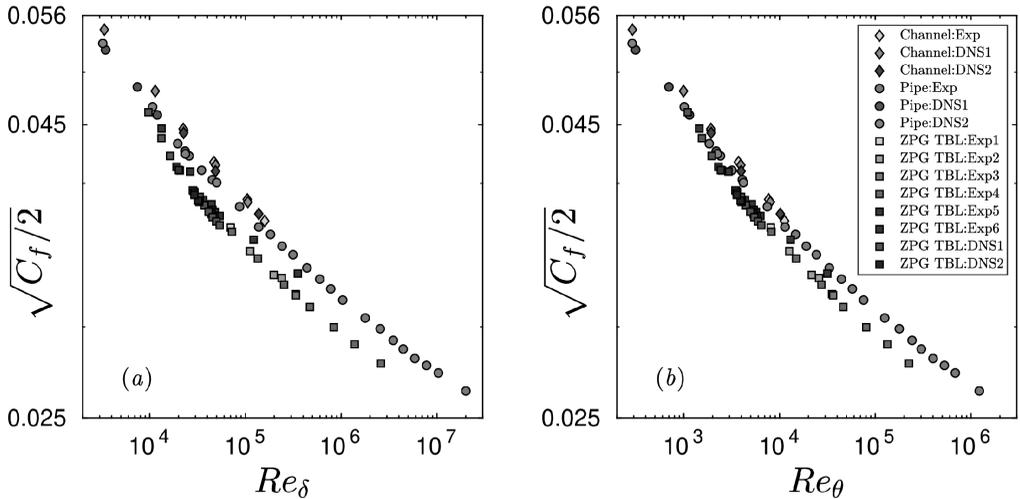}}% Images in 100% size
%  \centerline{\includegraphics[scale=0.33]{CfvsRetheta.eps}}% Images in 100% size
  \caption{Skin friction data of tables~\ref{table:channelpipe}~and~\ref{table:ZPG} plotted in the traditional $(\Rey,\cf)$ space. Note that $\sqtcfbytwodef$ and is plotted against two Reynolds numbers, $\Redelta$ in plot (\emph{a}) and $\Retheta$ in plot (\emph{b}). Symbol sizes are not indicative of the uncertainty in the data.}
\label{fig:CfRe}
\end{figure}

Figure~\ref{fig:CfRe} shows the skin friction data of tables~\ref{table:channelpipe}~and~\ref{table:ZPG} in the traditional space of $\Rey$ and $\cf$ variables; note that here $\uinf$ is the freestream velocity for ZPG TBLs and centreline velocity for pipes and channels. Two Reynolds numbers $\Redelta$ (figure~\ref{fig:CfRe}\emph{a}) and $\Retheta$ (figure~\ref{fig:CfRe}\emph{b}) are used. In this space, the data from different types of flows do not scale and collapse to a universal curve. In fact, the data appear to cluster around three distinct curves corresponding to the three types of flows under consideration. One reason for this lack of scaling could be the fundamental difference in the outer boundary conditions of these flows (see section~\ref{sec:generalflows}). Data in the $(\Retheta,\cf)$ space (figure~\ref{fig:CfRe}\emph{b}) show better clustering and reduced differences in the trends compared to the $(\Redelta,\cf)$ space (figure~\ref{fig:CfRe}\emph{a}). Even so, figure~\ref{fig:CfRe}(\emph{b}) shows that at high Reynolds numbers, the pipe flow data show a consistent shift of approximately $+7\%$ with respect to the ZPG TBL data. The typical skin friction measurement uncertainty for ZPG TBLs is $\pm 2.5\%$ in $\ut$ \citep{dixit2020PoF} and that for pipe flows is $\pm 0.5\%$ in $\ut$ \citep{dixit2021PoF}; significantly lower uncertainty in pipe flows is due to the accurate measurements of pressure drop along the length of the pipe that are used to infer skin friction \citep{dixit2021PoF}. Thus, the differences in the trends and the lack of scaling seen in figure~\ref{fig:CfRe} are well outside the measurement uncertainties and are therefore, genuine. Although channel flow data are not available at high Reynolds numbers, noticeable differences exist between channel and pipe flow data even at lower Reynolds numbers. In view of the discussion in section~\ref{subsec:dynamicallyconsistentscale}, this lack of collapse is not surprising. These observations underscore the fact that the traditional $(\Rey,\cf)$ space is not very useful towards universal scaling of skin friction in wall turbulence.

\section{Further analysis of the data: the $(\ltl,\utl)$ space}\label{sec:furtheranalysis}

Theory presented in section~\ref{sec:generalflows} shows that the asymptotic $-1/2$ power-law scaling of skin friction (\ref{eqn:asymplaw1}) and the corresponding finite-$\Rey$ model (\ref{eqn:finitelaw1}) must hold individually for any type of flow irrespective of the effects of flow geometry or outer boundary condition. We now demonstrate that the data from individual flow types indeed provide strong evidence in support of this expectation.  

\subsection{$M$-$\nu$ scaling for individual flows in the $(\ltl,\utl)$ space}\label{subsec:furtheranalysis-individualMnu}

\begin{figure}
  \centerline{\includegraphics[scale=0.475]{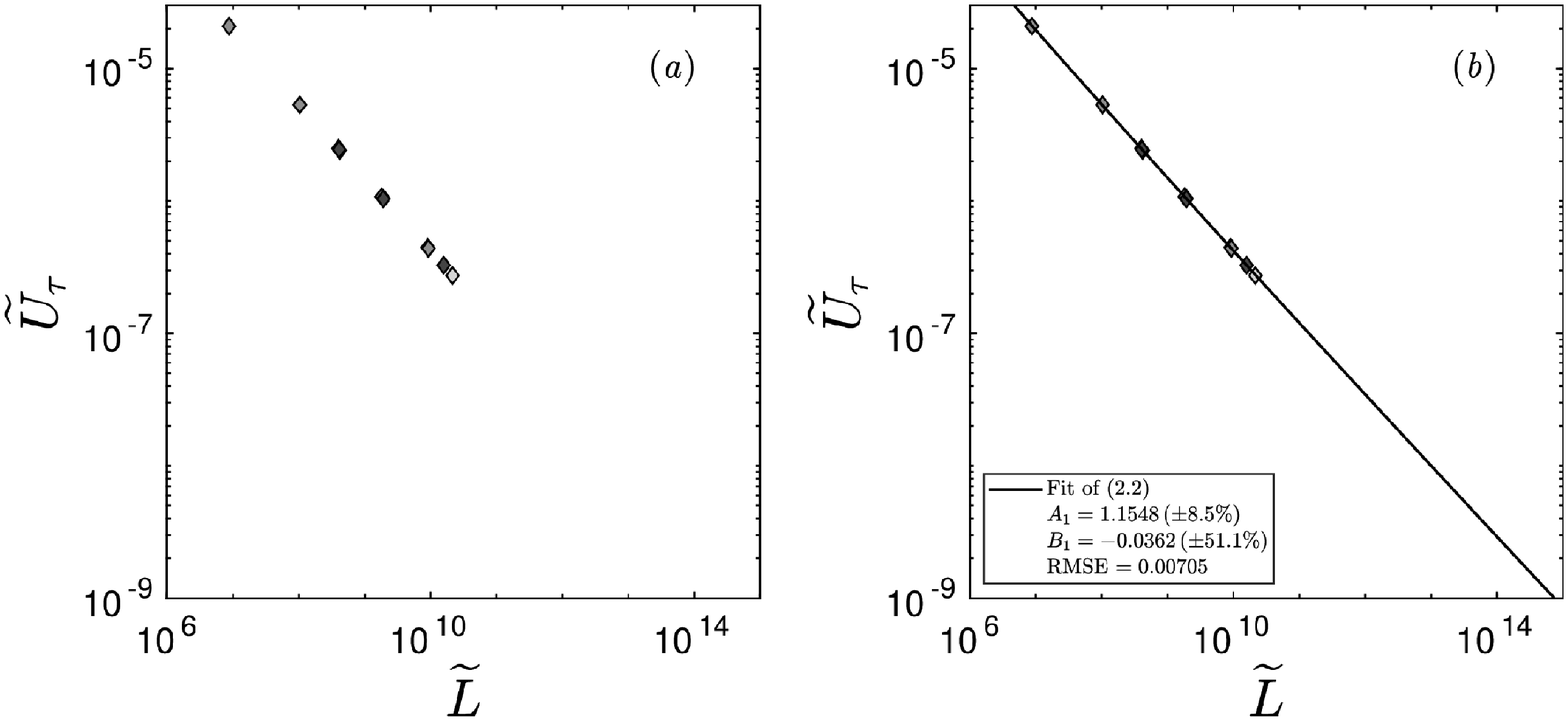}}% Images in 100% size
  \centerline{\hspace{5pt}\includegraphics[scale=0.475]{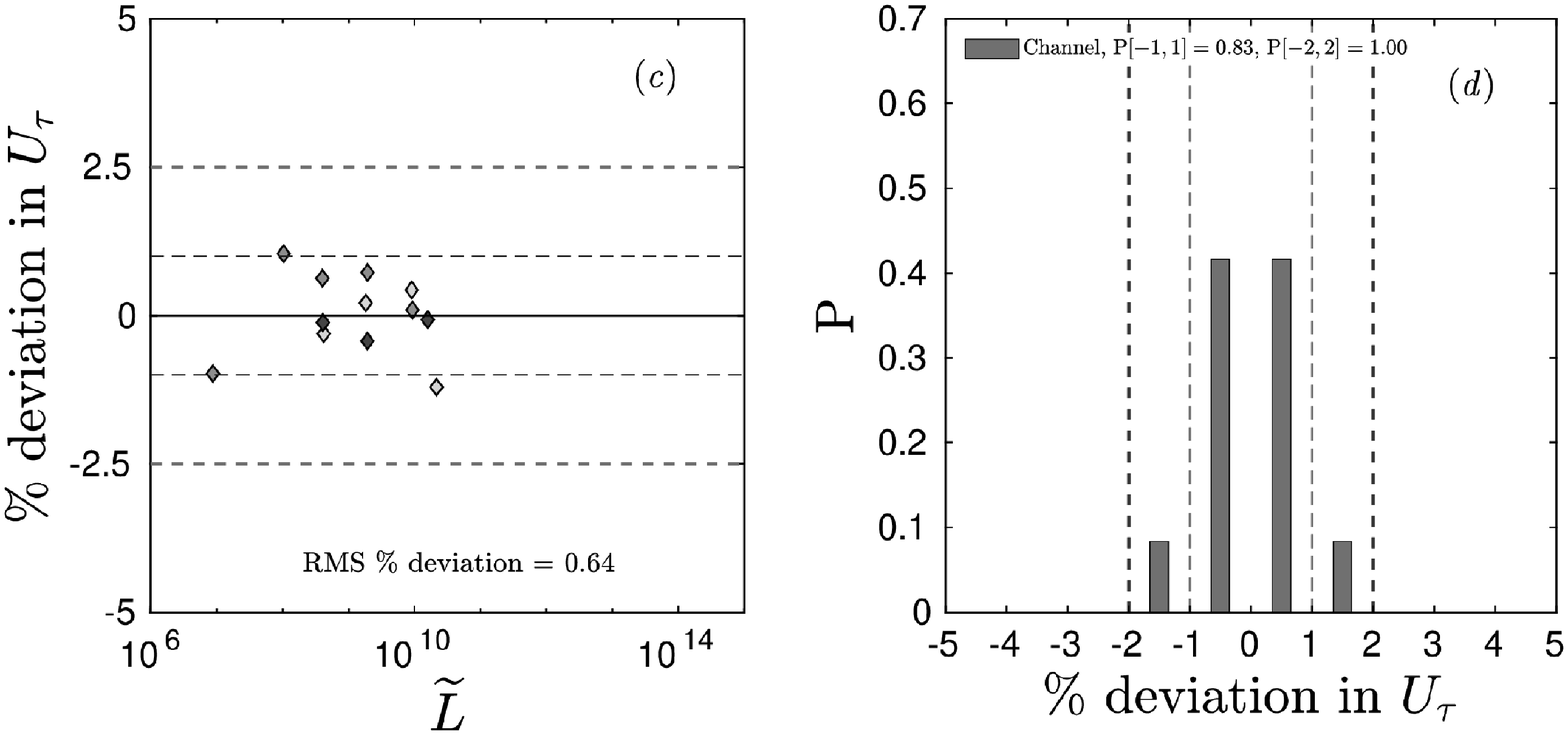}}% Images in 100% size
  \caption{Channel-flow skin friction data of table~\ref{table:channelpipe}. (\textit{a}) The $(\ltl,\utl)$ space. (\textit{b}) Least-squares fit of the finite-$\Rey$ model (\ref{eqn:finitelaw1}) to the data of plot (\textit{a}). Fitting constants are also shown along with the RMSE for the fit.(\textit{c}) Percentage deviations in the actual values of $\ut$ with respect to those computed using the finite-$\Rey$ model fitted in plot (\textit{b}). Thin dashed lines indicate the band of $[-1,1]$ and thick dashed lines indicate the band of $[-2.5,2.5]$. (\textit{d}) Probability histogram for the percentage deviations. Probability values for the two bands $[-1,1]$ (thin dashed lines) and $[-2,2]$ (thick dashed lines) are also shown. }
\label{fig:utlltlchannel}
\end{figure}

\begin{figure}
  \centerline{\includegraphics[scale=0.475]{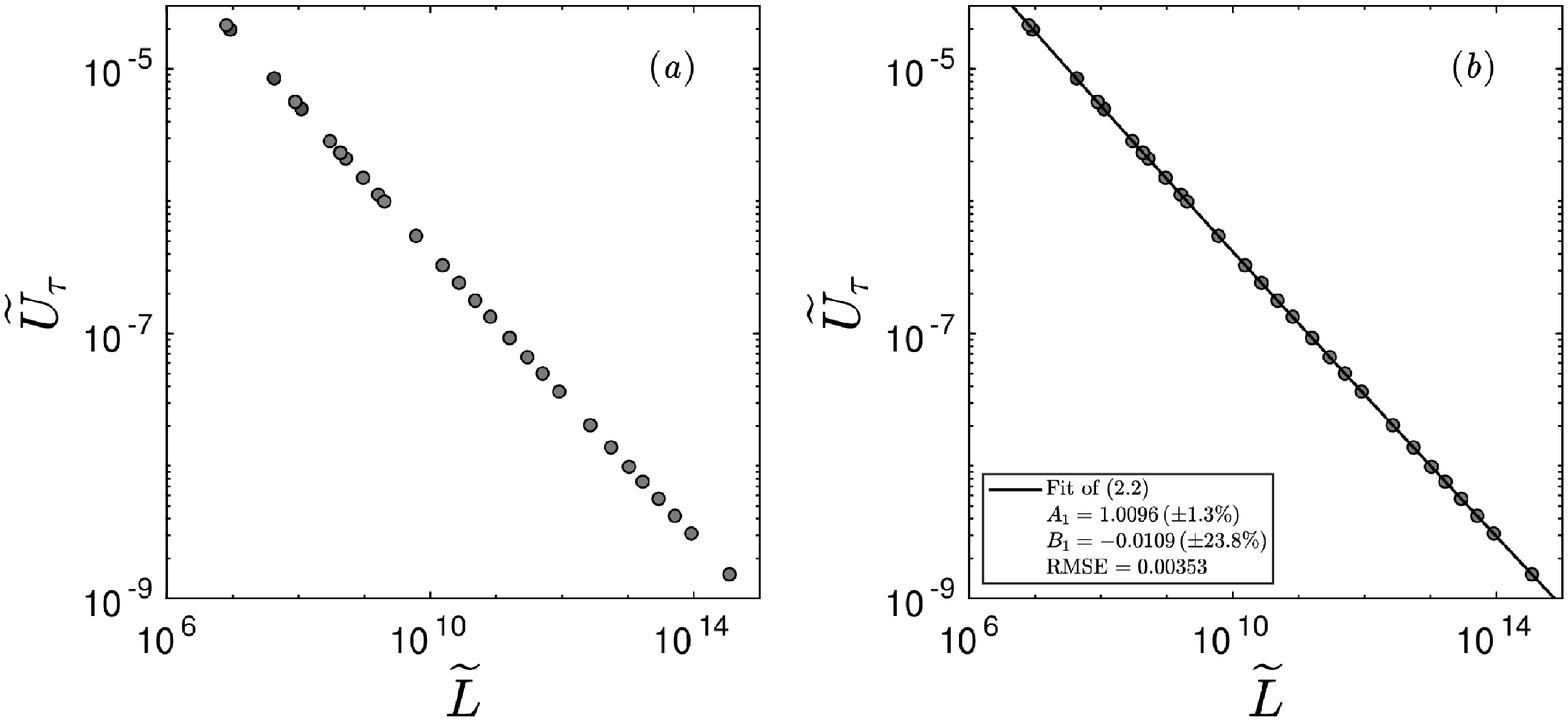}}% Images in 100% size
  \centerline{\hspace{5pt}\includegraphics[scale=0.475]{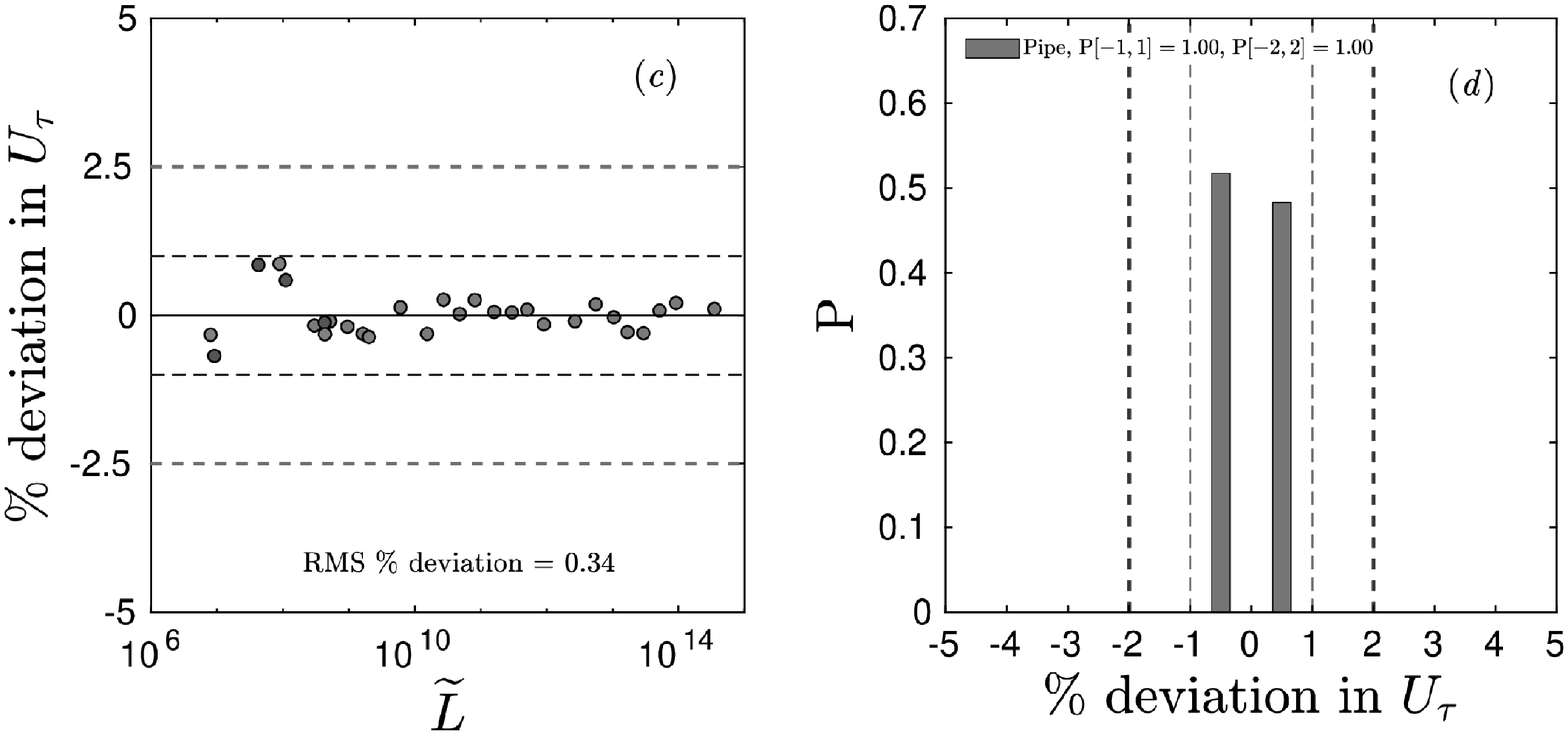}}% Images in 100% size
  \caption{Pipe-flow skin friction data of table~\ref{table:channelpipe}. Remaining details same as the caption of figure~\ref{fig:utlltlchannel}.}
\label{fig:utlltlpipe}
\end{figure}

\begin{figure}
  \centerline{\includegraphics[scale=0.475]{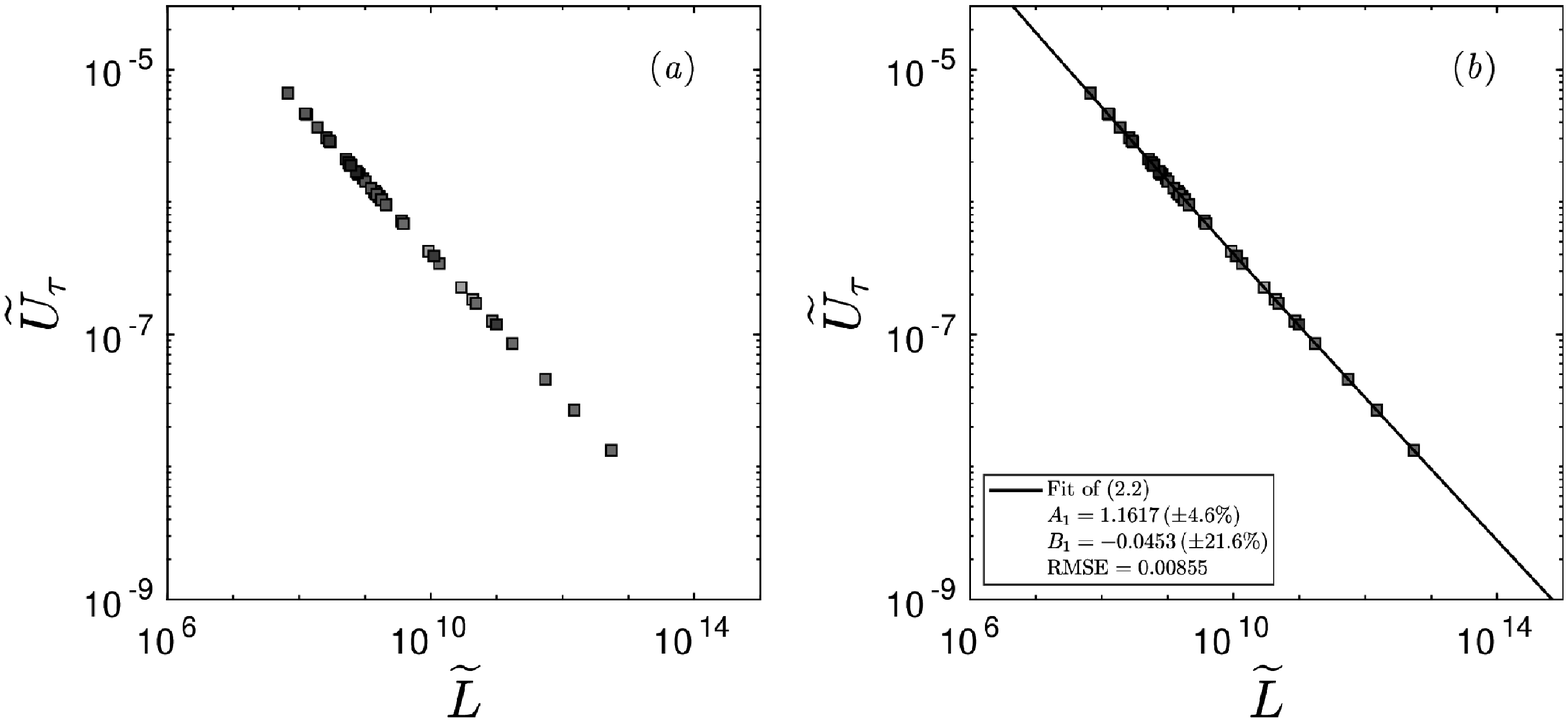}}% Images in 100% size
  \centerline{\hspace{5pt}\includegraphics[scale=0.475]{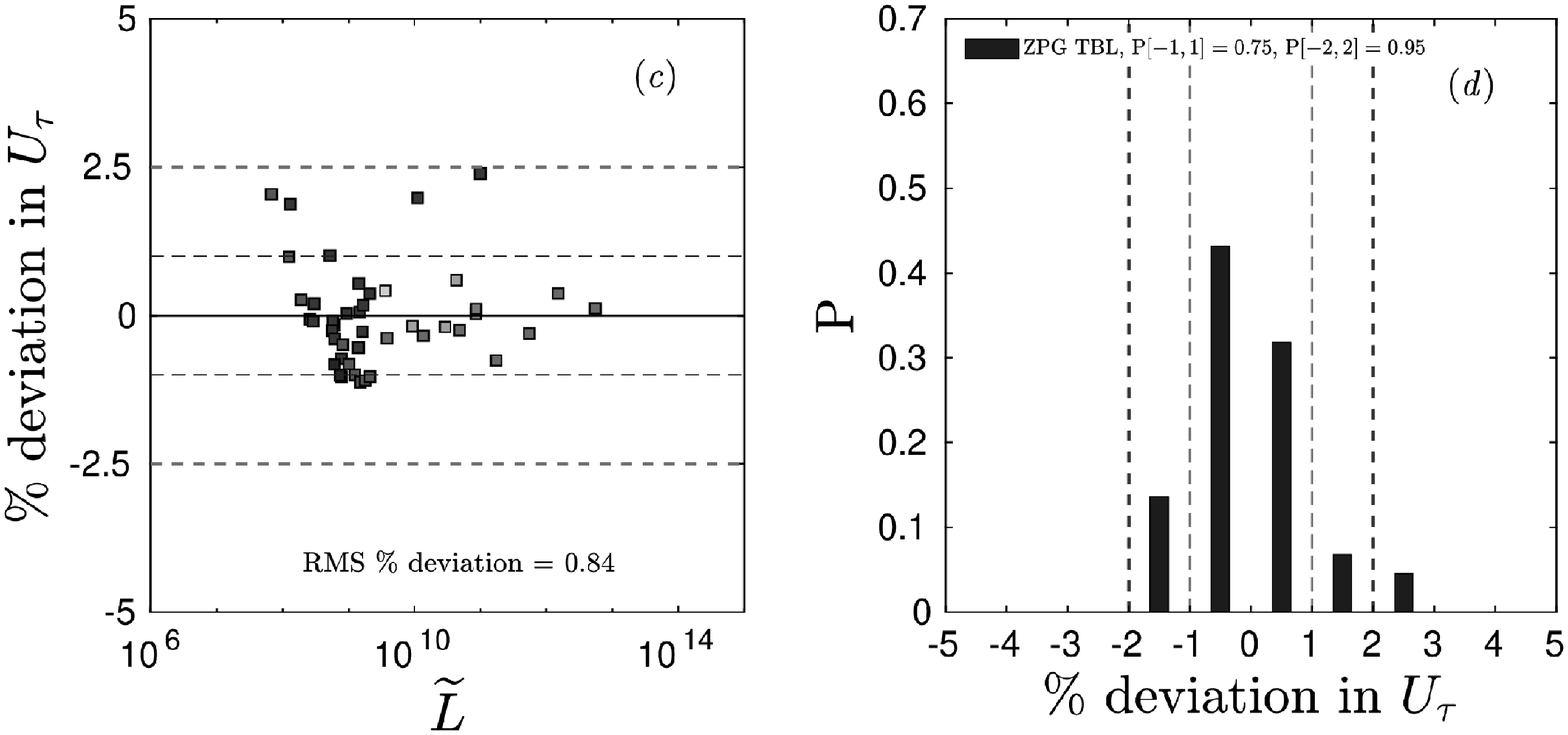}}% Images in 100% size
  \caption{ZPG TBL skin friction data of table~\ref{table:ZPG}. Remaining details same as the caption of figure~\ref{fig:utlltlchannel}.}
\label{fig:utlltlTBL}
\end{figure}

Figure~\ref{fig:utlltlchannel}(\emph{a}) shows the channel-flow data of table~\ref{table:channelpipe} plotted in the space of $\ltl$ and $\utl$ variables. Clearly, the data appear to line up along a single curve exhibiting scaling in this space as expected from the theory presented in section~\ref{sec:generalflows}. The equation of this curve is mathematically given by the finite-$\Rey$ model (\ref{eqn:finitelaw1}). A least-squares fit of this model to the data is shown in figure~\ref{fig:utlltlchannel}(\emph{b}). The values of model constants $A_{1}$ and $B_{1}$ in (\ref{eqn:finitelaw1}) after performing the fit, are also shown in figure~\ref{fig:utlltlchannel}(\emph{b}) along with the respective $95\%$ confidence intervals given parenthetically. Departure of the data from the fitted model is quantified by the root-mean-squared error (RMSE) and is also displayed in figure~\ref{fig:utlltlchannel}(\emph{b}) for ease of comprehension. In order to visualize how well the data scale in figure~\ref{fig:utlltlchannel}(\emph{b}), we compute the percentage deviation of the actual values of $\ut$ with respect to those obtained from the fit of the model and these deviations are shown in figure~\ref{fig:utlltlchannel}(\emph{c}). It is clear that almost all the percentage deviations are limited to the interval $[-1,1]$ which is at par with the measurement uncertainty of skin friction in channel flows \citep{schultz2013}; the RMS percentage deviation is $0.64\%$. Figure~\ref{fig:utlltlchannel}(\emph{d}) shows the probability histogram of the percentage deviations of figure~\ref{fig:utlltlchannel}(\emph{c}). The histogram is quite narrow and almost without any skew. This indicates that the finite-$\Rey$ model (\ref{eqn:finitelaw1}) shown in figure~\ref{fig:utlltlchannel}(\emph{b}) describes the channel-flow data quite well. Figures~\ref{fig:utlltlpipe} and \ref{fig:utlltlTBL} show the same processing of the pipe and ZPG TBL data respectively. The conclusions are largely the same. Somewhat increased scatter for the percentage $\ut$ deviations in ZPG TBLs (figure~\ref{fig:utlltlTBL}\emph{c}) and a weak skew in the corresponding probability histogram (figure~\ref{fig:utlltlTBL}\emph{d}) may be attributed to the inherent larger measurement uncertainty of $\ut$ ($\pm2.5\%$) in ZPG TBLs;  for pipes and channels this uncertainty is about $\pm1\%$ or so since there is privilege to infer the friction velocity from accurate pressure drop measurements due to the fully-developed character of internal flows.

On the whole, therefore, the results of figures~\ref{fig:utlltlchannel}, \ref{fig:utlltlpipe} and \ref{fig:utlltlTBL} provide compelling evidence in favour of the theory presented in section~\ref{sec:generalflows} and the individual applicability of (\ref{eqn:asymplaw1}) and (\ref{eqn:finitelaw1}) to internal (pipe and channel) as well as external (ZPG TBL) flows.

\subsection{Does $M$-$\nu$ scaling in $(\ltl,\utl)$ space hold for different flow types taken together?}\label{subsec:furtheranalysis-allMnu}

\begin{figure}
  \centerline{\includegraphics[scale=0.475]{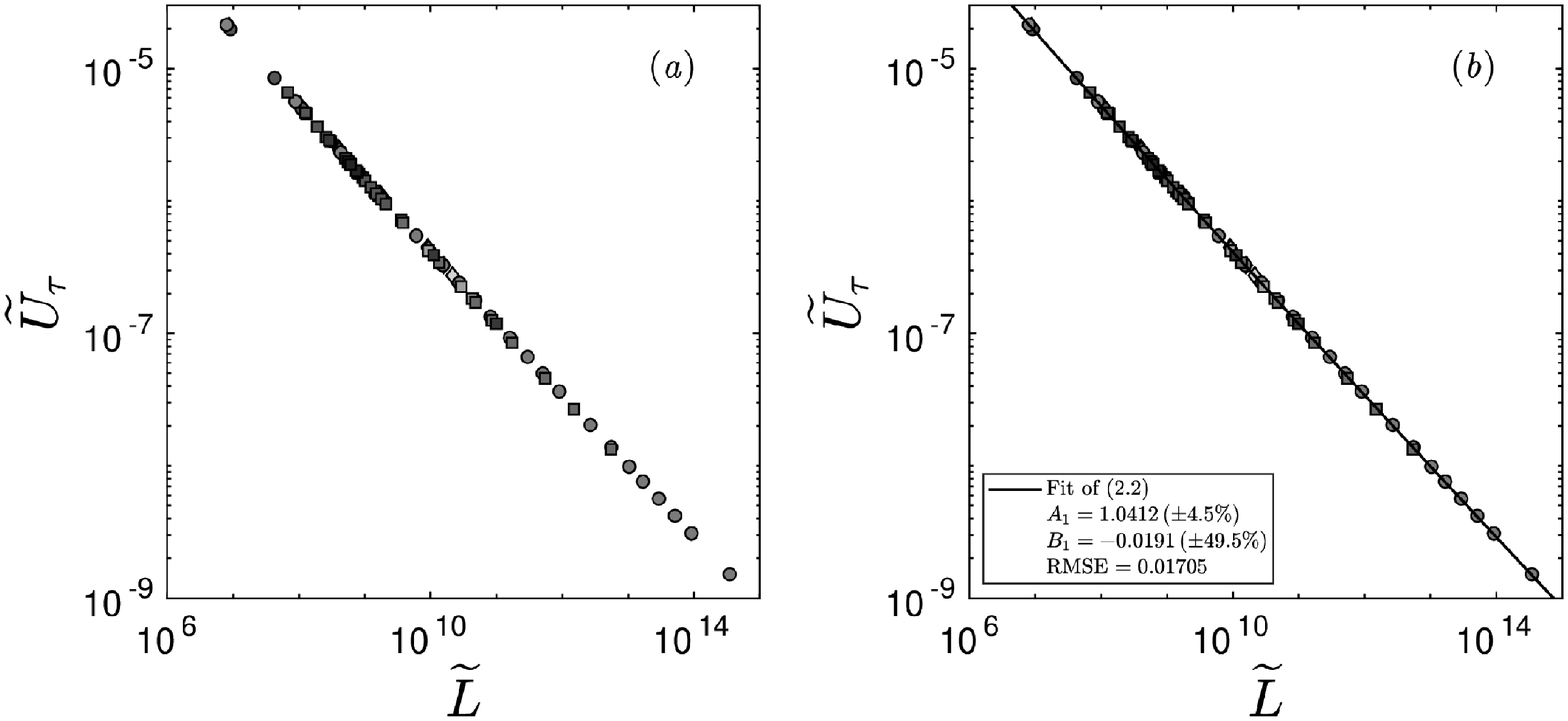}}% Images in 100% size
  \centerline{\hspace{5pt}\includegraphics[scale=0.475]{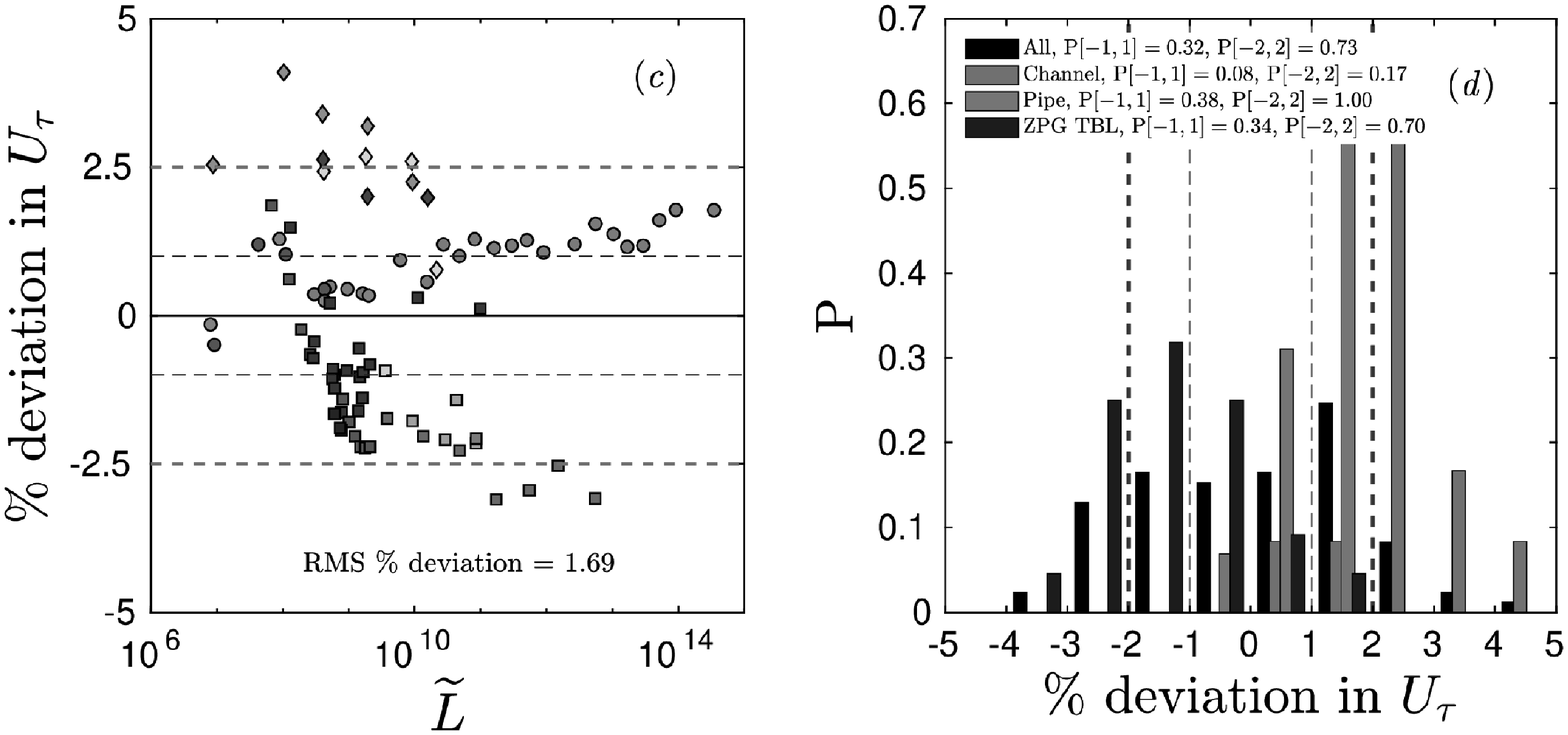}}% Images in 100% size
  \caption{Skin friction data of all the flows listed in tables~\ref{table:channelpipe}~and~\ref{table:ZPG}. Remaining details same as the caption of figure~\ref{fig:utlltlchannel}. Plot (\textit{d}) shows probability histograms for the percentage deviations for the individual flow types as well as for the complete data.}
\label{fig:utlltl}
\end{figure}

Figure~\ref{fig:utlltl} shows the same processing of the data as for figures~\ref{fig:utlltlchannel}-\ref{fig:utlltlTBL} but this time with data from all different flow types (tables~\ref{table:channelpipe} and \ref{table:ZPG}) put together. Figure~\ref{fig:utlltl}(\emph{b}) shows the fit of the finite-$\Rey$ model (\ref{eqn:finitelaw1}) performed for \emph{all} the data taken together. It is clear that the RMSE value for the fitted curve in figure~\ref{fig:utlltl}(\emph{b}) shows a two-fold increase compared to the RMSE values seen in figures~\ref{fig:utlltlchannel}-\ref{fig:utlltlTBL}(\emph{b}). Figure~\ref{fig:utlltl}(\emph{c}) shows that most of the deviations remain confined to the band of $\pm2.5\%$ in $\ut$ which is typical of skin friction measurement uncertainties in ZPG TBLs; the RMS percentage deviation is $1.69$. Specifically, the differences in the deviations of ZPG TBLs and pipe flow data at high Reynolds numbers are now confined to less than $5\%$ which is better as compared to the disagreement of $7\%$ noted earlier for the traditional $(\Retheta,\cf)$ space. This demonstrates that the $M$-$\nu$ scaling enables better collapse of \emph{all} the data from different flows in comparison to the traditional scaling. However, notwithstanding this, the ZPG TBL deviations in figure~\ref{fig:utlltl}(\emph{c}) systematically drift to the negative values with increasing Reynolds number. The opposite is true for the pipe flow data. This can be readily seen in figure~\ref{fig:utlltl}(\emph{d}) where the probability histograms of percentage $\ut$ deviations are biased to the right side for the pipe and channel flows, and to the left side for the ZPG TBLs; the probability histogram for all the data is therefore rather flat. Therefore, one may conclude that the data from different types of flows do not collapse onto each other and hence, onto the curve of the fitted model. The ZPG TBL data in figure~\ref{fig:utlltl}(\emph{b}) increasingly deviate downward with respect to the model curve with Reynolds number while the pipe (and channel) flow data continue to deviate upward. These persistent differences in the trends of different flow types, even after plotting in the $M$-$\nu$ scaling, appear to be related to the differences in the outer boundary condition amongst different flow types (see section~\ref{subsec:largescalestructures}). The asymptotic $-1/2$ power law (\ref{eqn:asymplaw1}) and the finite-$\Rey$ model (\ref{eqn:finitelaw1}) do not depend on the flow boundary conditions as shown by the theory in section~\ref{sec:generalflows} and demonstrated convincingly in section~\ref{subsec:furtheranalysis-individualMnu}. However, the values of the constants $A_{1}$ and $B_{1}$ in the finite-$\Rey$ model (\ref{eqn:finitelaw1}) - which describes the data trends of individual flows to an excellent accuracy - could certainly vary from one flow to the other depending on the boundary conditions. This variation is clear from figures~\ref{fig:utlltlchannel}(\emph{b}), \ref{fig:utlltlpipe}(\emph{b}) and \ref{fig:utlltlTBL}(\emph{b}). Therefore, absorbing the effects of the outer boundary condition into the framework of $M$-$\nu$ scaling appears to be the key to universal scaling of skin friction.
  
%\subsection{Shape factor correction for outer boundary condition: $M$-$\nu$-$G$ scaling in the $(\ltl',\utl')$ space}\label{subsec:analysis-MnuG}
\section{Universal $M$-$\nu$-$G$ scaling of skin friction in ZPG TBLs, pipes and channels}\label{sec:newscaling}

Discussion in the preceding shows that the differences in the outer boundary condition amongst various types of flows could become important if one seeks a universal description of skin friction and Reynolds number in the $(\ltl,\utl)$ space. In what follows, we begin by considering why the outer boundary conditions could at all become important. The answer to this question lies in the fact that the large-scale structures contribute significantly to the mean skin friction and these structures are strongly influenced by the outer boundary conditions. Next, we shall discuss how to actually account for the outer boundary condition effects in the scaling and this will be followed by presentation of the new scaling and assessing its efficacy.

\subsection{Large-scale structures and outer boundary condition effects}\label{subsec:largescalestructures}

Historically, the large-scale structures in the outer layer of wall-bounded turbulent flows were conceived to be the horseshoe-shaped or hairpin-shaped eddies anchored to the wall or the `attached' eddies of turbulence \citep{perry1982}. The largest size of an attached eddy, which might well get physically disconnected from the wall at a later stage during its lifetime, was limited by the flow thickness (boundary layer thickness or pipe radius or channel half-height). However, \cite{kim1999} reported energetic peak in the spectrum of streamwise velocity fluctuation $u'$ in pipe flows, at long wavelengths of the order of $12R$, $R$ being the pipe radius; these outer-scaled motions were termed as `very large scale motions' (VLSMs). The occurrence of energetic motions at such long wavelengths hinted at structural organisation. \cite{zhou1999} reported that such structural `organisation' of the hairpin vortices in channel flows takes the form of streamwise coherent vortex packets (called large scale motions or LSMs) wherein all the hairpins belonging to a packet travel at the same speed in the downstream direction. Further work on the hairpin packets \citep{adrian2000,christensen2001,adrian2007} showed that the typical streamwise extent of a packet is approximately $3\delta$, $\delta$ being the boundary layer thickness, and a train of packets aligned one after the other was considered as a possible explanation of the occurrence of VLSMs and hence the spectral peaks at wavelengths longer than $\delta$. 

The near-wall influences of the large-scale outer layer motions was a topic of controversy in the early seventies and eighties. For example, while \cite{rao1971} reported that the frequency of the near-wall bursting events in a ZPG TBL flow scales on the boundary layer thickness $\delta$, \cite{blackwelder1983} contested this conclusion. In the recent times however, more affirmative and concrete evidence of such influences came from the experimental work of \cite{hutchins2007a} in ZPG TBLs. Using spectral maps of $u'$ obtained over a range of Reynolds numbers from the HRNBLWT facility at the University of Melbourne, they were able to show that the energetic outer-layer structures in ZPG TBLs have long wavelengths of the order of $10\delta$ and these were termed as `superstructures' \citep{hutchins2007a}. Due to the spanwise meandering character of the superstructures, their signature in the $u'$ spectrum was shown to appear at shorter wavelengths of the order of $6\delta$ \citep{hutchins2007b,abbassi2017}. These superstructures give rise to the outer energy site of (located in the log region of ZPG TBLs) TKE which grows stronger with Reynolds number and exerts a progressively increasing influence on the $u'$ statistics in the near-wall region. This influence can be seen as the slow (logarithmic) increase, with Reynolds number, of the peak value of the inner-scaled streamwise turbulence intensity at $\yplus\approx15$ \citep{hutchins2007b}. These ideas were further pursued in a series of papers \citep[][to name a few]{hutchins2007b,mathis2009,mathis2011a} that resulted in a model that could predict, to a good accuracy, the near-wall turbulence statistics given a single streamwise velocity timeseries measurement in the log layer of a ZPG TBL flow \citep{marusic2010Sci,adrian2010,mathis2011b}. 

While these earlier works studied the `footprinting' (superposition) and `amplitude modulation' effects of the large-scale motions on the near-wall velocity fluctuations (primarily $u'$), more recent studies have focussed on the contributions to the mean skin friction \citep{hwang2017,fan2019}, skin friction statistics \citep{agostini2019} and various strategies of skin friction reduction and flow control \citep{abbassi2017,kim2017LEBU}. These studies mostly use DNS data of fully-developed channel flows, although some flow control studies use DNS of ZPG TBLs. Insights into the mechanisms of how exactly the large-scale structures influence the skin friction, have been gained. For example, \cite{agostini2019} suggest that the large-scale motions alter the shape of the instantaneous velocity profiles thus altering the instantaneous turbulence production in the buffer layer of a channel flow. This causes the small-scale motions to amplify or attenuate in an indirect manner rather than experiencing direct amplitude modulation under the influence of the large scales. Notwithstanding the differences in the mechanisms, all the studies unanimously conclude that a significant fraction of the mean skin friction is attributed to the large-scale motions and increases with Reynolds number \citep{hwang2013,giovenatti2016,hwang2017,agostini2019,fan2019}. 

In order to understand the possible connection between the large-scale structures and the outer boundary condition, it is important to characterize these in pipes, channels and ZPG TBLs at matched Reynolds number $\Retau$. To our knowledge, a single systematic study of this sort is by \cite{monty2009} which provides several interesting insights. The spectral structure of the above-mentioned flows has been experimentally studied using careful single hotwire measurements at matched $\Retau$ ($\approx 3000$) and hotwire spatial resolution $l_{+}=l\ut/\nu=30$. The study has revealed important structural differences between internal (pipe and channel) and external (ZPG TBL) flows. The $u'$ spectra in the ZPG TBL show a spectral peak around the superstructure wavelength of about $6\delta$ in the log region, but this peak quickly attenuates to the LSM wavelength of $3\delta$ with the spectrum remaining largely unimodal as one moves away from the log region into the outer layer (figure~3\emph{b} in Monty \etal). The pipe and channel spectra in the outer layer however, exhibit strong bimodal character; the LSM mode remains fixed at $3\delta$ ($\delta$ is the boundary layer thickness, pipe radius and channel half-height) but the longer wavelength (VLSM) mode shows scale growth reaching $\approx14$ to $20\delta$ wavelengths (figures~3\emph{d} and 3\emph{f} in Monty \etal). Further, in the near-wall region where all the three flows are supposed to behave universally, it is found that the spectral structure is not universal. Even at $\yplus\approx15$, where the motions are supposed to be dominated by the universal inner (wall-scaled) viscous cycle, the ZPG TBL spectrum shows higher spectral density at viscous-scaled wavelength of $1000$ units than the pipe and channel spectra (figure~4\emph{a} in Monty \etal). Since the streamwise intensity profiles collapse for all these flows (figure~1\emph{b} in Monty \etal) due to matched $\Retau$ and $l_{+}$, the above differences in the spectra imply redistribution of the TKE from shorter to longer wavelengths in pipes and channels as compared to the ZPG TBLs (evident in figure~4\emph{a} in Monty \etal). Although the intensity profiles collapse, the mean velocity profiles however, show the increasing degree of fullness (or decreasing strength of the wake component) in the order ZPG TBL, pipe and channel as expected (figures~1\emph{a} and 2\emph{a} in Monty \etal). Monty \etal~ finally comment that ``\ldots it is clear that the conditions in pipes/channels must permit the very large modes to persist further from the wall than in boundary layers (in which they are largely constrained to the log region) \ldots Also, far from the wall $(z/\delta\geq0.6)$, the geometrical freedom of the boundary layer is highlighted as energy ultimately decays to zero at the edge of the boundary layer, while the internal flows remain turbulent through the core\ldots ". The `conditions' and the `geometrical freedom' that Monty \etal~refer to, are precisely the outer boundary condition effects that are proposed here to introduce non-universality into the scaling of skin friction. A very interesting recent numerical study by \cite{kwon2021} in channel flows, shows that the VLSMs, although present in the further part of the log region, essentially bear a strong link to the outer layer (or the outer boundary condition); the autonomous log layer structure does not show evidence of VLSMs. The mechanism for the influence of large-scale structures on the mean skin friction therefore appears to be the effect of the outer boundary condition in terms of deciding the relative dominance of the large-scale structures in the outer layer of internal and external flows and hence, their varying contributions to the mean skin friction. These structural differences are consistent with the differences in the wake strength of the mean velocity profile for these flows as mentioned earlier in this section. Therefore, an indirect way to account for these structural effects could be to account for the shape of the mean velocity profile in these flows.

\subsection{Clauser shape factor $G$}\label{subsec:G}

One major manifestation of the outer boundary condition is the fullness of the mean velocity profile in a given flow. It is a standard practice to express the mean velocity profiles of ZPG TBLs, pipes and channels in terms of the wall-wake formulation of \cite{coles1956}. Further, it is well-known that for the same $\Retau$, the wake factor - a parameter indicative of the strength of the wake component in the mean velocity profile - decreases in the order ZPG TBL, pipe and channel \citep{monty2009}. Since the wake factor is indicative of the mean velocity deficit compared to the freestream or centreline velocity, a lower value of wake factor implies lower strength of the wake component, lower mean velocity deficit and hence, larger fullness of the mean velocity profile. Therefore, for a given $\Retau$, the fullness of the mean velocity profile increases in the order ZPG TBL, pipe and channel \citep{monty2009}.

An objective integral measure of the fullness of the velocity profile is its shape factor. For laminar boundary layers, only one velocity scale ($\uinf$) and one length scale ($L$) are sufficient to achieve complete self-similarity of velocity profiles due to the single-layer scaling description. In such cases, the conventional shape factor $H$ (ratio of displacement to momentum thickness) is a constant for a given self-similar flow regardless of the skin friction or Reynolds number. For TBL flows, however, the situation is complicated due to the problem of multiple scales. The inner or wall layer is governed by the viscous scales $\ut$ and $\nu/\ut$ whereas the outer layer is better described in terms of the defect scaling - the departure of the mean velocity from its freestream value scales on $\ut$ and the wall-normal distances scale on $L$. One then talks of the self-similarity of inner-scaled and defect-scaled mean velocity profiles. Due to this, as pointed out by \cite{clauser1956}, the conventional shape factor $H$ (ratio of displacement to momentum thickness) in ZPG TBLs varies with the skin friction coefficient $\cf$ or the flow Reynolds number $\Rey$ even though the defect profiles exhibit self-similarity. Hence, $H$ is not a very useful measure of the fullness of the TBL mean velocity profiles because self-similarity demands that the shape factor should essentially be constant for a given self-similar TBL flow. To resolve this problem, \cite{clauser1956} went on to define an integral thickness 
\begin{equation}
\Delta\coloneqq\int_{0}^{\infty}\left(\frac{U-\uinf}{\ut}\right)\textrm{d}y,\label{eqn:clauserdelta}
\end{equation} 
\noindent and a new shape factor
\begin{equation}
G\coloneqq\int_{0}^{\infty}\left(\frac{U-\uinf}{\ut}\right)^{2}\textrm{d}\left(\frac{y}{\Delta}\right),\label{eqn:clauserG}
\end{equation} 
\noindent based on the self-similarity of \emph{defect} velocity profiles. Thus, for self-similar (in outer or defect coordinates) TBL flows, $G$ is essentially a constant, and the parameters $H$, $G$ and $\cf$ are related \citep{clauser1956} by 
\begin{equation}
H=\left(1-G\sqrt{\cf/2}\right)^{-1}.\label{eqn:HGCf}
\end{equation} 
Since the flows of the present interest (ZPG TBLs, pipes and channels) are turbulent, the Clauser shape factor $G$ is a more appropriate measure to account for the differences in the fullness of the mean velocity profiles than the conventional shape factor $H$.
 
In the present context we note that, $G\approx 6.8$  for ZPG TBLs at high Reynolds numbers \citep{clauser1956}. Further, since the velocity profile fullness increases (i.e. defect decreases) in the order ZPG TBL, pipe and channel \citep[see][]{monty2009}, the asymptotic values of $G$ are expected to be $G_{\textrm{\emph{ZPG TBL}}}>G_{\textrm{\emph{pipe}}}>G_{\textrm{\emph{channel}}}$. If one takes the asymptotic value of $G$ for ZPG TBLs as an arbitrary, yet useful, reference i.e. $G_{\textrm{\emph{ref}}}=6.8$, then the difference between the fullness of a mean velocity profile (in any flow) with respect to the reference fullness of a ZPG TBL profile could be quantified by $(G-G_{\textrm{\emph{ref}}})/G_{\textrm{\emph{ref}}}$ or simply by the ratio $\GbyGref$. Furthermore, rearrangement of (\ref{eqn:HGCf}) shows that
\begin{equation}
\frac{G}{G_{\textrm{\emph{ref}}}}=\frac{\uinf}{G_{\textrm{\emph{ref}}}\ut}\left(\frac{H-1}{H}\right)=\frac{\uinftl}{G_{\textrm{\emph{ref}}}\utl}\left(\frac{H-1}{H}\right),\label{eqn:GbyGrefexpr}
\end{equation} 
\noindent where $\uinftl=\uinftlexp$. Note that, $\uinftl$, $H$ and $G_{\textrm{\emph{ref}}}$ are known parameters from the measured or computed mean velocity distribution, and thus, $\GbyGref$ itself is a function of $\ut$ or $\utl$.  

\subsection{Universal, semi-empirical $M$-$\nu$-$G$ scaling and finite-$\Rey$ model: the $(\ltl',\utl')$ space}\label{subsec:MnuGscaling}

\begin{figure}
  \centerline{\includegraphics[scale=0.475]{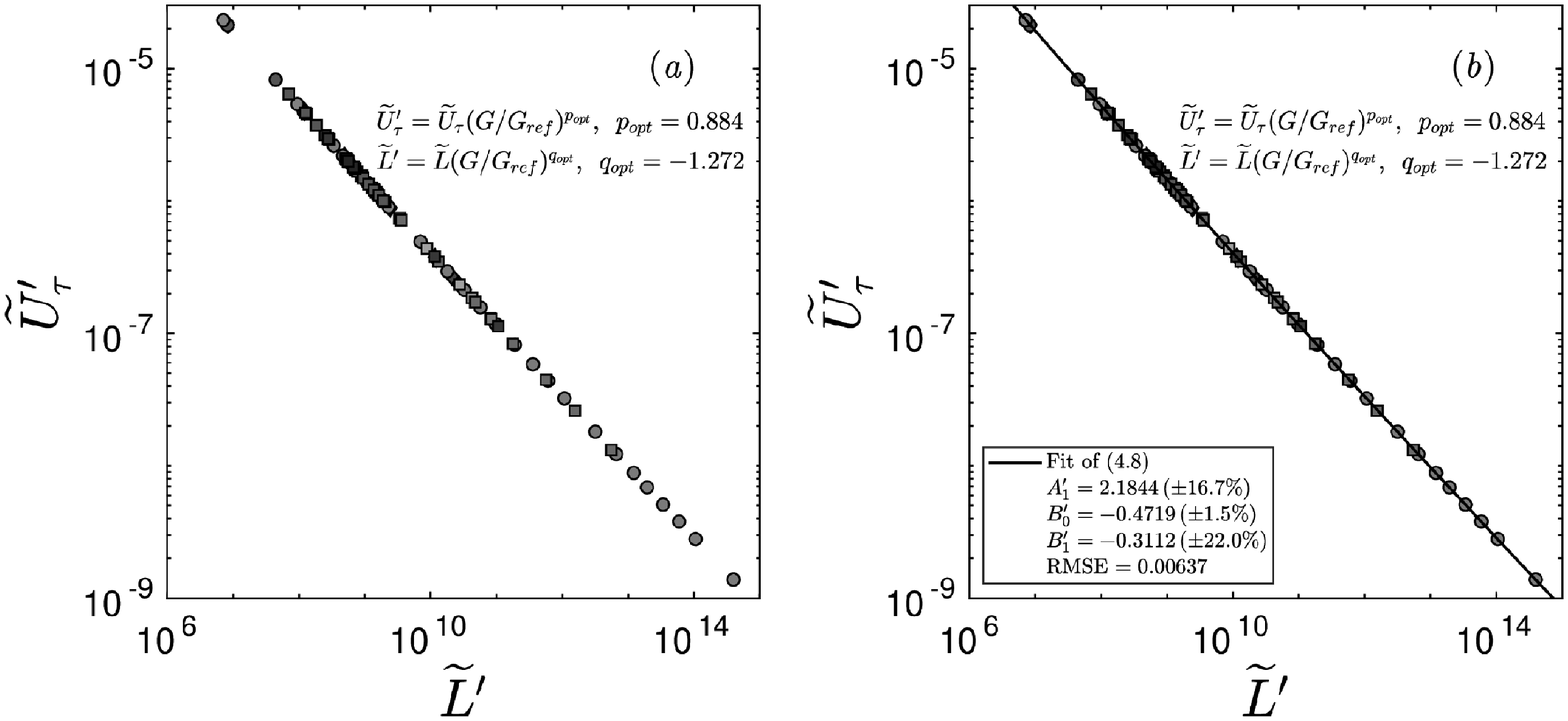}}% Images in 100% size
%  \centerline{\hspace{2pt}\includegraphics[scale=0.33]{UtpvsLtpmodel.eps}}% Images in 100% size
  \centerline{\hspace{5pt}\includegraphics[scale=0.475]{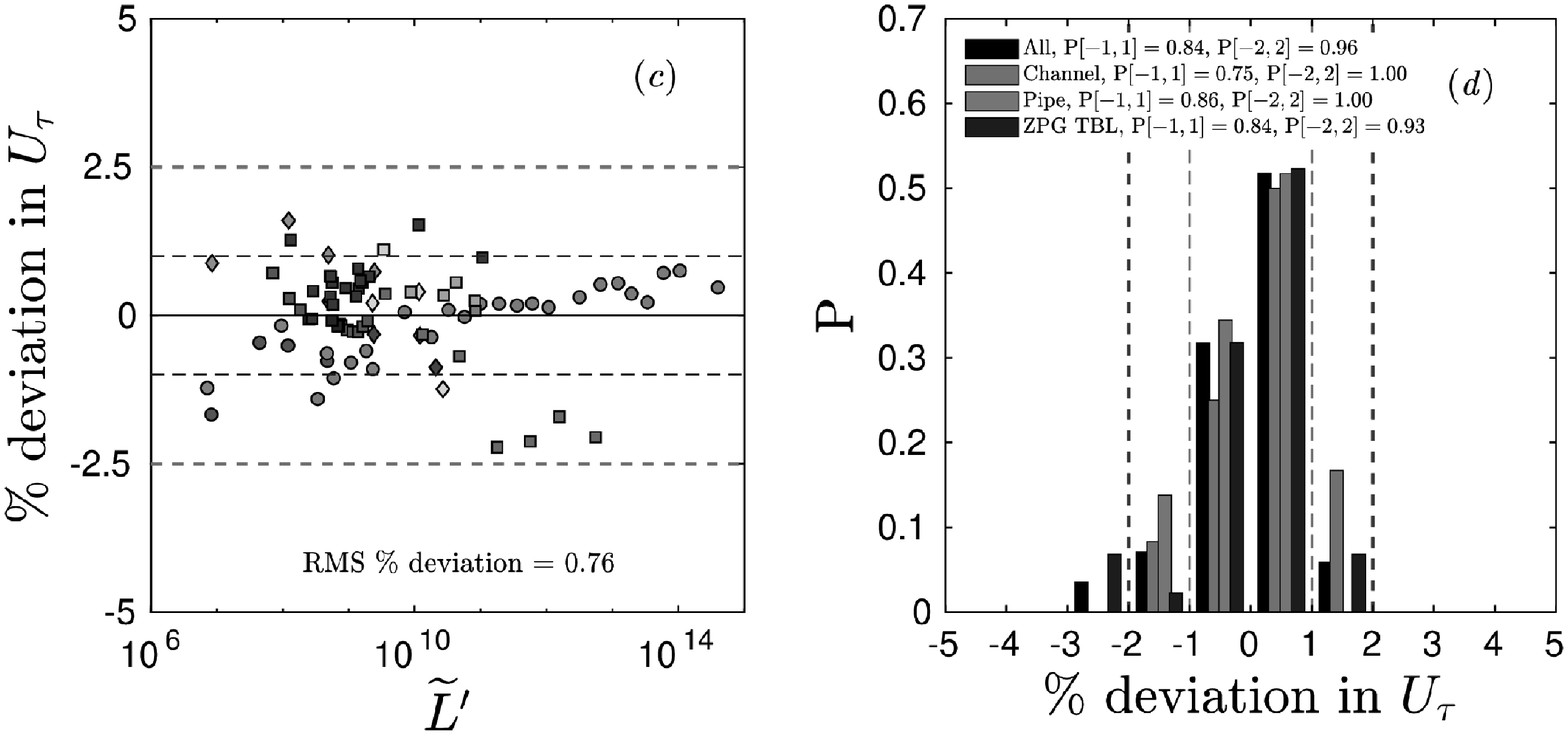}}% Images in 100% size
  \caption{(\textit{a}) Skin friction data of tables~\ref{table:channelpipe}~and~\ref{table:ZPG} in the $(\ltl',\utl')$ space. (\textit{b}) Least-squares fit of the finite-$\Rey$ model (\ref{eqn:finitelaw2}) to the data of plot (\textit{a}). Fitting constants are also shown along with the RMSE for the fit.(\textit{c}) Percentage deviations in the actual values of $\ut$ with respect to those computed using the finite-$\Rey$ model fitted in plot (\textit{b}). Thin dashed lines indicate the band of $[-1,1]$ and thick dashed lines indicate the band of $[-2.5,2.5]$. (\textit{d}) Probability histograms for the percentage deviations for the individual flow types as well as for the complete data. Probability values for the two bands $[-1,1]$ (thin dashed lines) and $[-2,2]$ (thick dashed lines) are also shown.}
\label{fig:utl'ltl'}
\end{figure}

With the preceding discussion in sections~\ref{subsec:largescalestructures} and \ref{subsec:G} as a plausible physical basis, we attempt to introduce an empirical correction factor based on $\GbyGref$, within the framework of $M$-$\nu$ scaling i.e. the $(\ltl,\utl)$ space. The expectation is that this $M$-$\nu$-$G$ scaling would lead to better collapse of the data from all the flows than the $M$-$\nu$ scaling alone. Towards this, we propose an empirical transformation of the variables $\ltl$ and $\utl$ using correction factors that are functions of the ratio $\GbyGref$. Specifically, we assume the correction factors to have a power-law functional form and define our new \emph{shape-factor-corrected} variables as
\begin{eqnarray}
\utl'&\coloneqq&\utl(G/G_{\textrm{\emph{ref}}})^{p},\label{eqn:utl'}\\
\ltl'&\coloneqq&\ltl(G/G_{\textrm{\emph{ref}}})^{q},\label{eqn:ltl'}
\end{eqnarray}
\noindent where $p$ and $q$ ($p\neq q$, in general) are empirical constants. Counterparts of the asymptotic law (\ref{eqn:asymplaw1}) and finite-$\Rey$ model (\ref{eqn:finitelaw1}), in terms of the shape-factor-corrected variables $\ltl'$ and $\utl'$, would then be
\begin{eqnarray}
\utl'&\sim&\ltl'^{B'_{0}},\label{eqn:asymplaw2}\\
\utl'&=&\frac{A'_{1}}{\ln\ltl'}\ltl'^{\left[B'_{0}+\frac{B'_{1}}{\sqrt{\ln\ltl'}}\right]},\label{eqn:finitelaw2}
\end{eqnarray}
\noindent where $A'_{1}$, $B'_{0}$ and $B'_{1}$ are constants. Note that if one considers only ZPG TBLs, then \mbox{$\GbyGref\approx 1$} so that, (\ref{eqn:asymplaw2}) and (\ref{eqn:finitelaw2}) reduce to (\ref{eqn:asymplaw1}) and (\ref{eqn:finitelaw1}) respectively as expected. The constant $B'_{0}$ in (\ref{eqn:asymplaw2}) and (\ref{eqn:finitelaw2}) is now expected to be different from $-1/2$ because of the transformation of $\utl$ and $\ltl$ to $\utl'$ and $\ltl'$ respectively. It is re-emphasized that the proposed corrections, although guided by the data and the physical understanding outlined before, are purely empirical in nature, and the contention is that the shape-factor-corrected finite-$\Rey$ model (\ref{eqn:finitelaw2}), in the $M$-$\nu$-$G$ scaling, universally describes all the data from different types of flows to an approximation significantly better than (\ref{eqn:finitelaw1}). 

In order to complete the universal finite-$\Rey$ model (\ref{eqn:finitelaw2}), one needs to determine the values of five parameters namely $p$, $q$, $A'_{1}$, $B'_{0}$ and $B'_{1}$. Towards this, we use the following approach involving simple optimization and model curve fitting. First, we arbitrarily select the values for $p$ and $q$, and compute $\ltl'$ and $\utl'$ for the data. Next, we perform a nonlinear least-squares fit (using the MATLAB function \texttt{nlinfit}) of the shape-factor-corrected finite-$\Rey$ model (\ref{eqn:finitelaw2}) to the data, and obtain the root-mean-squared error (RMSE) of the data with respect to the fitted curve. Finally, we systematically and independently vary the values of $p$ and $q$ each over the range $-10$ to $10$, and for each pair of values $(p,q)$, we repeat the above procedure and monitor the RMSE value. The pair $(p,q)$ exhibiting the minimum RMSE for the fitted model is considered as the optimized pair $(p_{\textrm{\emph{opt}}},q_{\textrm{\emph{opt}}})$, and the values of $A'_{1}$, $B'_{0}$ and $B'_{1}$ corresponding to this fitting of the model are finalized for further analysis. 

Figure~\ref{fig:utl'ltl'}(\emph{a}) shows all the data plotted in the space of shape-factor-corrected variables $(\ltl',\utl')$; the corresponding fit of (\ref{eqn:finitelaw2}) is shown in figure~\ref{fig:utl'ltl'}(\emph{b}). Note that, the optimization and curve-fitting process mentioned in the preceding paragraph goes back and forth between figures~\ref{fig:utl'ltl'}(\emph{a}) and \ref{fig:utl'ltl'}(\emph{b}). The optimum values of $p$ and $q$ in (\ref{eqn:utl'}) and (\ref{eqn:ltl'}) turn out to be $p_{\textrm{\emph{opt}}}=0.884$ and $q_{\textrm{\emph{opt}}}=-1.272$ as shown in figure~\ref{fig:utl'ltl'}(\emph{a}) as well as figure~\ref{fig:utl'ltl'}(\emph{b}). Also, the values of the constants in (\ref{eqn:finitelaw2}) turn out to be $A_{1}'=2.1844$, $B_{0}'=-0.4719$ and $B_{1}'=-0.3112$ as shown in figure~\ref{fig:utl'ltl'}(\emph{b}). A careful look at figures~\ref{fig:utlltl}(\emph{a}) and \ref{fig:utl'ltl'}(\emph{a}) indicates discernible improvement in data collapse when the shape factor correction is included i.e. switching over from the $M$-$\nu$ scaling to the $M$-$\nu$-$G$ scaling. Figures~\ref{fig:utlltl}(\emph{b}) and \ref{fig:utl'ltl'}(\emph{b}) quantify this improvement; RMSE with reference to the respective finite-$\Rey$ model significantly improves by almost an order of magnitude from $0.01705$ (figure~\ref{fig:utlltl}\emph{b}) to $0.00637$ (figure~\ref{fig:utl'ltl'}\emph{b}). Also, it is interesting to note that $B_{0}'=-0.4719$ with the inclusion of shape factor effect (see figure~\ref{fig:utl'ltl'}\emph{b}). This value is close to the asymptotic value of $-0.5$ expected from the theory presented in section~\ref{sec:generalflows}; it must be remembered, however, that the theory applies to each type of flow individually irrespective of the boundary conditions and does not include the shape factor correction which is essential to collapse the data from flows with different boundary conditions. This suggests that, while the shape factor correction could be considered as a small perturbation from the asymptotic behaviour for each flow, it is crucial to improve the scaling of the data from flows with different boundary conditions. Figure~\ref{fig:utl'ltl'}(\emph{c}) demonstrates that the shape factor correction dramatically reduces the differences in the trends noted earlier in figure~\ref{fig:utlltl}(\emph{c}). This is clear from the reduction in RMS percentage deviation in $\ut$ by a factor of two from $1.69$ (figure~\ref{fig:utlltl}\emph{c}) to $0.76$ (figure~\ref{fig:utl'ltl'}\emph{c}). Figure~\ref{fig:utl'ltl'}(\emph{d}) reflects this dramatic improvement in the data collapse in terms of clustering of the probability histograms for all the types of flows around the percentage $\ut$ deviation value of zero; figure~\ref{fig:utl'ltl'}(\emph{d}) is to be contrasted with figure~\ref{fig:utlltl}(\emph{d}). shows this Thus, the contention that the shape factor effect could play a key role in the universal scaling of skin friction in wall turbulence is strongly supported by the improved scaling of the data in the $(\ltl',\utl')$ space i.e. the $M$-$\nu$-$G$ scaling.

\subsection{Converting from $(\ltl',\utl')$ space to three-dimensional $(\ltl,\GbyGref,\utl)$ space}

Equation (\ref{eqn:finitelaw2}) universally (for all flows) relates the variables $\ltl'$ and $\utl'$ in the $M$-$\nu$-$G$ scaling according to the general functional form
\begin{equation}
\utl'=F_{2}(\ltl'),\label{eqn:finitelaw2generalform}
\end{equation}
\noindent so that, $\utl'$ is a function of a single variable $\ltl'$. Equation~(\ref{eqn:finitelaw2generalform}) may be converted back to the original variables $\ltl$ and $\utl$ along with the third variable $\GbyGref$ that amounts to the shape factor correction. Expanding (\ref{eqn:finitelaw2}) using the definitions (\ref{eqn:utl'}) and (\ref{eqn:ltl'}), and rearrangement yields
\begin{equation}
\utl=\frac{A'_{1}}{\ln\left[\ltl(\GbyGref)^{q}\right]}\left[\ltl(\GbyGref)^{q}\right]^{\left[B'_{0}+\frac{B'_{1}}{\sqrt{\ln\left[\ltl(\GbyGref)^{q}\right]}}\right]}(\GbyGref)^{-p}.\label{eqn:finitelaw2-3D}
\end{equation}
\noindent Equation (\ref{eqn:finitelaw2-3D}) shows that $\utl$ is a function of two variables namely $\ltl$ and $\GbyGref$, wherein $\GbyGref$ in turn depends on $\utl$ according to (\ref{eqn:GbyGrefexpr}). Therefore, the universal finite-$\Rey$ model for skin friction (\ref{eqn:finitelaw2}) translates to an implicit (in $\utl$) expression of the general form
\begin{equation}
\utl=F_{3}(\ltl,\GbyGref),\label{eqn:finitelaw2generalform-3D}
\end{equation}
\noindent where the functional form of $F_{3}$ is given by the right side of (\ref{eqn:finitelaw2-3D}). 

\begin{figure}
  \centerline{\includegraphics[scale=0.5]{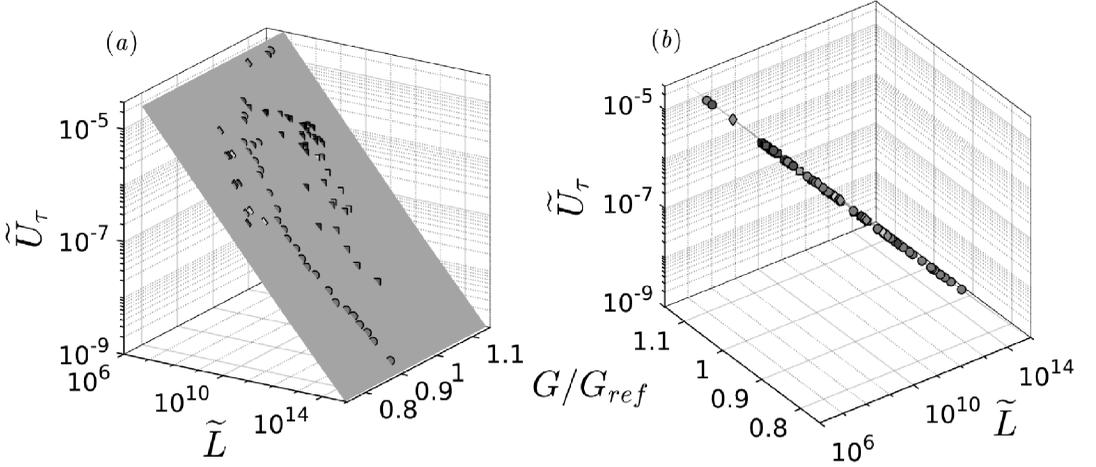}}% Images in 100% size
  \caption{Skin friction data in the three-dimensional space of $\ltl$, $\GbyGref$ and $\utl$ variables as per (\ref{eqn:finitelaw2-3D}) and (\ref{eqn:finitelaw2generalform}). Shading represents the surface given by (\ref{eqn:finitelaw2-3D}) using $p=p_{\textrm{\emph{opt}}}=0.884$, $q=q_{\textrm{\emph{opt}}}=-1.272$, $A_{1}'=2.1844$, $B_{0}'=-0.4719$ and $B_{1}'=-0.3112$ as obtained earlier in figures~\ref{fig:utl'ltl'}(\emph{a}) and \ref{fig:utl'ltl'}(\emph{b}). Plot (\emph{b}) shows an almost along-the-surface view of the same data as in plot (\emph{a}) indicating tight collapse of the data points onto the surface given by (\ref{eqn:finitelaw2-3D}).}
\label{fig:utlltlGbyGref3D}
\end{figure}

Figure~\ref{fig:utlltlGbyGref3D} shows all the data plotted in the three-dimensional space $(\ltl,\GbyGref,\utl)$. Also plotted is the surface given by (\ref{eqn:finitelaw2-3D}) with values of all the empirical constants same as those obtained in the previous section. It is clear that the data points conform very well to the surface implying universal scaling according to (\ref{eqn:finitelaw2}) or (\ref{eqn:finitelaw2-3D}). 

\section{Conclusion}\label{sec:conclusion}

Scaling of mean skin friction in different types of turbulent wall-bounded flows has been investigated in this work. Specifically, we explore the possibility of a universal scaling of skin friction in ZPG TBLs, pipes and channels. These flow types are the most widely studied in the literature and represent canonical flow archetypes of wall turbulence. The main points addressed in this work may be summarized as follows:

\begin{enumerate}
\item Based on the process of conversion of the streamwise mean flow kinetic energy into the kinetic energy of turbulence by the largest eddies of the flow, a new derivation (section~\ref{sec:generalflows}) of the asymptotic skin friction law (\ref{eqn:asymplaw1}) is presented. This process of kinetic energy conversion is independent of the boundary conditions of the flow. Therefore, the asymptotic law (\ref{eqn:asymplaw1}) and the finite-$\Rey$ model (\ref{eqn:finitelaw1}) based on that, are both independent of the boundary conditions of the flow. Therefore, (\ref{eqn:asymplaw1}) and (\ref{eqn:finitelaw1}) individually apply to ZPG TBLs, pipes and channels which are the focus of the present paper. Several high-quality experimental and DNS data sets, from the literature on ZPG TBLs, pipes and channels, are analysed to investigate the occurrence of the scaling predicted by the new theory. The data from each type of flow provide strong support in favour of (\ref{eqn:asymplaw1}) and (\ref{eqn:finitelaw1}) holding very well in individual flows.

\item The velocity scale $M/\nu$ for skin friction - $M$ is the kinematic momentum rate of the flow in the streamwise-wall-normal plane and $\nu$ is the fluid kinematic viscosity - is seen to emerge from the governing dynamical equations. Thus, the $M$-$\nu$ scaling embodied in (\ref{eqn:asymplaw1}) and (\ref{eqn:finitelaw1}) is dynamically consistent. In contrast, the freestream(bulk) velocity scale $\uinf$($U_{b}$) used in the traditional ZPG TBL(pipe/channel) relations, while easy-to-measure(practically relevant), is more of a boundary condition than the governing dynamics.    

\item It is known that, the large-scale structures in the outer layer of ZPG TBLs, pipes and channels contribute significantly to mean skin friction, and this contribution progressively increases with Reynolds number. It has also been suggested \citep{monty2009} that differences in the outer boundary condition amongst different flow types lead to differences in the relative dominance of large-scale structures in these flows as is evident from the spectral redistribution of TKE. These facts suggest that outer boundary conditions have a role to play in the universal scaling of mean skin friction across different types of wall-bounded turbulent flows. The most striking manifestation of the distinctive effect of outer boundary condition in different types of flows is the difference in the shape (fullness) of mean velocity profile. Clauser's shape factor $G$ is proposed as a parameter that could account for the differences in skin friction scaling due to differences in outer boundary condition across different types of flows. With the asymptotic value of $G$ for ZPG TBLs as a reference ($G_{\textrm{\emph{ref}}}=6.8$), the ratio $\GbyGref$ is proposed as a measure of disparity between the outer boundary condition effects in different flows. Empirical inclusion of a correction factor based on $\GbyGref$ into the framework of $M$-$\nu$ scaling is shown to lead to a new, universal $M$-$\nu$-$G$ scaling and the new, universal, finite-$\Rey$ model given by (\ref{eqn:finitelaw2}). This new scaling collapses data from ZPG TBLs, pipes and channels remarkably well onto a single, universal curve given by (\ref{eqn:finitelaw2}) in the space of shape-factor-corrected variables $(\ltl',\utl')$ or a single, universal surface given by (\ref{eqn:finitelaw2-3D}) in the three-dimensional space $(\ltl,\GbyGref,\utl)$.

%The derivation A new derivation of the Using the integral momentum equation for ZPG TBLs, \cite{dixit2020PoF} earlier showed that the variation of skin friction (or friction velocity $\ut$) with Reynolds number may be effectively `scaled' if one uses a velocity scale $M/\nu$ derived from the TBL kinematic momentum rate $M$ and fluid kinematic viscosity $\nu$. In the limit of infinite Reynolds number, they showed that the asymptotic friction law takes the form $\utl\sim\ltl^{-1/2}$. A finite-$\Rey$ model for skin friction (\ref{eqn:finitelaw1}) for ZPG TBLs was also developed therein.
%\item In this work, we have shown that 

\end{enumerate}

We thank Professor I. Marusic for the original ZPG TBL data sets from the Melbourne experiments and Professor M. Schultz for the original channel flow data sets from the Maryland experiments. We also thank Professor B.~J.~McKeon and Dr.~R.~Chin for providing links to their pipe flow experimental and DNS data sets. We also thank all the other researchers for making their experimental or DNS data (listed in table~\ref{table:refslinks}) available for free download. SAD thanks Professor O. N. Ramesh for discussions regarding the importance of shape factor in wall-bounded turbulent flows. All authors gratefully acknowledge the continued support by the Director, IITM, Pune. We dedicate this work to the memory of the Late Professor R. Narasimha.

Declaration of Interests. The authors report no conflict of interest.

Data Availability. All the data used in this work are available in the literature and at the links cited in table~\ref{table:refslinks}. No new data have been generated.  

\appendix
\section{Data from the literature used in this work}\label{appA}

This appendix lists various parameters for experimental and DNS data sets used in the present study. Table~\ref{table:channelpipe} lists the parameters for the channel and pipe flow data sets whereas table~\ref{table:ZPG} lists them for the ZPG TBL data sets. The references for the data and their sources (digitized from a plot in the reference paper or provided by the authors or downloadable from a link) are listed in table~\ref{table:refslinks}.
\begin{table}
  \begin{center}
\def~{\hphantom{0}}
  \begin{tabular}{ccccccccc}%{lllllllll}%
      Data set  & $\ut$  & $\uinf$ & $L$ & $M$ & $\utl$ & $\ltl$ & $H$ & $\GbyGref$ \\
      code  & $(\textrm{ms}^{-1})$  & $(\textrm{ms}^{-1})$ & $(\textrm{m})$ & $(\textrm{m}^{3}\textrm{s}^{-2})$ &  &  & & \\[6pt]
      Channel:Exp & $0.0750$ & $1.687$ & $0.0127$ & $0.0292$ &$2.4120\times10^{-6}$ & $4.1857\times10^{8}$ & $1.356$ & $0.869$\\
			  & $0.1453$ & $3.481$ & $0.0127$ & $0.1267$ &$1.0778\times10^{-6}$ & $1.8152\times10^{9}$ & $1.309$ & $0.832$\\
			  & $0.2996$ & $7.734$ & $0.0127$ & $0.6278$ &$4.4681\times10^{-7}$ & $9.0598\times10^{9}$ & $1.278$ & $0.826$\\
			  & $0.4400$ & $11.861$ & $0.0127$ & $1.5167$ &$2.7393\times10^{-7}$ & $2.1519\times10^{10}$ & $1.269$ & $0.840$\\[3pt]
	Channel:DNS1 & $0.0544$ & $1.000$ & $0.0504$ & $0.0390$ &$2.0933\times10^{-5}$ & $8.7420\times10^{6}$ & $1.617$ & $1.030$\\
	            & $0.0481$ & $1.000$ & $0.1714$ & $0.1356$ &$5.3253\times10^{-6}$ & $1.0328\times10^{8}$ & $1.402$ & $0.877$\\
	            & $0.0446$ & $1.000$ & $0.3358$ & $0.2682$ &$2.4955\times10^{-6}$ & $4.0033\times10^{8}$ & $1.350$ & $0.855$\\
	            & $0.0415$ & $1.000$ & $	0.7310$ & $0.5945$ &$1.0469\times10^{-6}$ & $1.9314\times10^{9}$ & $1.305$ & $0.828$\\
	            & $0.0385$ & $1.000$ & $	1.5880$ & $1.3152$ &$4.3942\times10^{-7}$ & $9.2826\times10^{9}$ & $1.271$ & $0.815$\\[3pt]
	Channel:DNS2 & $0.0415$ & $1.103$ & $1.0000$ & $1.0127$ &$3.2772\times10^{-7}$ & $1.5825\times{10}$ & $1.259$ & $0.803$\\
	            & $0.0459$ & $1.119	$ & $1.0001$ & $1.0172$ &$1.0373\times10^{-6}$ & $1.9232\times10^{9}$ & $1.307$ & $0.842$\\
	            & $0.0500$ & $1.130	$ & $0.9995$ & $1.0205$ &$2.4511\times10^{-6}$ & $4.0798\times10^{8}$ & $1.349$ & $0.860$\\[6pt]
     Pipe:Exp	  & $0.2089^{\dagger}$ & $4.822$ & $0.0647$ & $1.1649$ &$2.8470\times10^{-6}$ & $2.9890\times10^{8}$ & $1.366$ & $0.910$\\
                 & $0.2683^{\dagger}$ & $6.347$ & $0.0647$ & $2.0311$ &$2.1012\times10^{-6}$ & $5.1902\times10^{8}$ & $1.353$ & $0.907$\\
                 & $0.3455^{\dagger}$ & $8.413$ & $0.0647$ & $3.5870$ &$1.5047\times10^{-6}$ & $9.5054\times10^{8}$ & $1.340$ & $0.908$\\
                 & $0.4320$ & $10.721$ & $0.0644$ & $5.9031$ &$1.1206\times10^{-6}$ & $1.6225\times10^{9}$ & $1.330$ & $0.906$\\
                 & $0.7919$ & $20.740$ & $0.0646$ & $22.2844$ &$5.4696\times10^{-7}$ & $6.0712\times10^{9}$ & $1.301$ & $0.892$\\
                 & $0.4183$ & $11.414$ & $0.0639$ & $6.7548$ &$3.2803\times10^{-7}$ & $1.5396\times10^{10}$ & $1.283$ & $0.884$\\
                 & $0.5437$ & $15.058$ & $0.0645$ & $11.9454$	&$2.4188\times10^{-7}$ & $2.7264\times10^{10}$ & $1.271$ & $0.869$\\
                 & $0.7035$ & $19.946$ & $0.0645$ & $21.0737$ &$1.7792\times10^{-7}$ & $4.7821\times10^{10}$ & $1.264$ & $0.871$\\
                 & $0.9003$ & $25.967$ & $0.0645$ & $35.8917$ &$1.3382\times10^{-7}$ & $8.1286\times10^{10}$ & $1.256$ & $0.864$\\
                 & $0.2423$ & $7.177$ & $0.0645$ & $2.7571$ &$9.2777\times10^{-8}$ & $1.5952\times10^{11}$ & $1.249$ & $0.870$\\
                 & $0.3230$ & $9.785$ & $0.0645$ & $5.1447$ &$6.6329\times10^{-8}$ & $2.9714\times10^{11}$ & $1.241$ & $0.866$\\
                 & $0.4136$ & $12.776$ & $0.0644$ & $8.7703$ &$4.9918\times10^{-8}$ & $5.0414\times10^{11}$ & $1.235$ & $0.864$\\
                 & $0.5411$ & $17.092$ & $0.0646$ & $15.8127$ &$3.6435\times10^{-8}$ & $9.0123\times10^{11}$ & $1.231$ & $0.872$\\
                 & $0.4721$ & $15.457$ & $0.0645$ & $12.9921$ &$2.0345\times10^{-8}$ & $2.6718\times10^{12}$ & $1.222$ & $0.876$\\
                 & $0.1759$ & $5.884$ & $0.0646$ & $1.8924$ &$1.3824\times10^{-8}$ & $5.5271\times10^{12}$ & $1.216$ & $0.872$\\
                 & $0.2358$ & $8.075$ & $0.0646$ & $3.5686$ &$9.8598\times10^{-9}$ & $1.0356\times10^{13}$ & $1.213$ & $0.885$\\
                 & $0.2147$ & $7.480$ & $0.0644$ & $3.0613$ &$7.6111\times10^{-9}$ & $1.6736\times10^{13}$ & $1.211$ & $0.891$\\
                 & $0.2782$ & $9.879$ & $0.0645$ & $5.3496$ &$5.6476\times10^{-9}$ & $2.9267\times10^{13}$ & $1.206$ & $0.890$\\
                 & $0.3652$ & $13.150$ & $0.0644$ & $9.4836$ &$4.1915\times10^{-9}$ & $5.1578\times10^{13}$ & $1.203$ & $0.894$\\
                 & $0.4821$ & $17.626$ & $0.0644$ & $17.1143$ &$3.0905\times10^{-9}$ & $9.1645\times10^{13}$ & $1.199$ & $0.894$\\
                 & $0.9127$ & $34.590$ & $0.0645$ & $66.9222$ &$1.5189\times10^{-9}$ & $3.4819\times10^{14}$ & $1.193$ & $0.901$\\[3pt]
     Pipe:DNS1	  & $0.0523$ & $1.000$ & $0.0520$ & $0.0396$ &$1.9812\times10^{-5}$ & $9.1387\times10^{6}$ & $1.628$ & $1.085$\\
                 & $0.0485$ & $1.000$ & $0.1116$ & $0.0857$ &$8.4897\times10^{-6}$ & $4.2548\times10^{7}$ & $1.469$ & $0.967$\\
                 & $0.0459$ & $1.000$ & $0.1798$ & $0.1383$ &$4.9774\times10^{-6}$ & $1.1051\times10^{8}$ & $1.416$ & $0.942$\\
                 & $0.0427$ & $1.000$ & $0.3511$ & $0.2737$ &$2.3389\times10^{-6}$ & $4.2714\times10^{8}$ & $1.367$ & $0.926$\\[3pt]
     Pipe:DNS2	  & $0.5295$ & $10.000$ & $0.0049$ & $0.3706$ & $2.1435\times10^{-5}$ & $7.9988\times10^{6}$ & $1.652$ & $1.096$\\
                 & $0.4663$ & $10.000$ & $0.0161$ & $1.2404$ & $5.6394\times10^{-6}$ & $8.8707\times10^{7}$ & $1.431$ & $0.942$\\
                 & $0.4243$ & $10.000$ & $0.0354$ & $2.7427$ & $2.3204\times10^{-6}$ & $4.3178\times10^{8}$ & $1.373$ & $0.942$\\
                 & $0.4006$ & $10.000$ & $0.0750$ & $6.0390$ & $9.9511\times10^{-7}$ & $2.0131\times10^{9}$ & $1.323$ & $0.897$
  \end{tabular}
  \caption{Experimental and DNS data of fully-developed channel and pipe flows from the literature. See table \ref{table:refslinks} for the references and data sources.}
  \label{table:channelpipe}
  \end{center}
\end{table}

\begin{table}
  \begin{center}
\def~{\hphantom{0}}
  \begin{tabular}{ccccccccc}%{lllllllll}%
      Data set  & $\ut$  & $\uinf$ & $L$ & $M$ & $\utl$ & $\ltl$ & $H$ & $\GbyGref$ \\
      code  & $(\textrm{ms}^{-1})$  & $(\textrm{ms}^{-1})$ & $(\textrm{m})$ & $(\textrm{m}^{3}\textrm{s}^{-2})$ &  &  & & \\[6pt]
ZPG TBL:Exp1	& $0.4575$ & $12.500$ & $0.0853	$ & $9.7116$ & $7.1579\times10^{-7}$ & $3.5878\times10^{9}$ & $1.350$ & $1.042$\\[3pt]
ZPG TBL:Exp2	& $0.7068$ & $20.239$ & $0.0838$ & $25.2919$ & $4.2274\times10^{-7}$ & $9.2575\times10^{9}$ & $1.329$ & $1.042$\\
	& $0.6824$ & $20.497$ & $0.1473$ & $46.1912$ & $2.2576\times10^{-7}$ & $2.9126\times10^{10}$ & $1.309$ & $1.044$\\
	& $0.6607$ & $19.968$ & $0.1833$ & $55.0135$ & $1.8353\times10^{-7}$ & $4.3183\times10^{10}$ & $1.295$ & $1.013$\\
	& $0.6379$ & $19.910$ & $0.2564$ & $77.1017$ & $1.2589\times10^{-7}$ & $8.5372\times10^{10}$ & $1.291$ & $1.035$\\[3pt]
ZPG TBL:Exp3	& $0.6073$ & $19.000$ & $0.2677$ & $72.8688$ & $1.2584\times10^{-7}$ & $8.5568\times10^{10}$ & $1.286$ & $1.023$\\[3pt]
ZPG TBL:Exp4	& $0.3296$ & $9.080$ & $0.0262$ & $1.5846$ & $6.8558\times10^{-7}$ & $3.8245\times10^{9}$ & $1.351$ & $1.053$\\
	& $0.3169$ & $9.210$ & $0.0273$ & $1.7252$ & $3.4314\times10^{-7}$ & $1.3508\times10^{10}$ & $1.313$ & $1.020$\\
	& $0.3033$ & $9.290$ & $0.0281$ & $1.8248$ & $1.7143\times10^{-7}$ & $4.8187\times10^{10}$ & $1.287$ & $1.005$\\
	& $0.2913$ & $9.330$ & $0.0265$ & $1.7960$ & $8.5086\times10^{-8}$ & $1.7297\times10^{11}$ & $1.262$ & $0.978$\\
	& $0.2836$ & $9.460$ & $0.0251$ & $1.7584$ & $4.5744\times10^{-8}$ & $5.4787\times10^{11}$ & $1.248$ & $0.973$\\
	& $0.2753$ & $9.500$ & $0.0240$ & $1.6999$ & $2.6735\times10^{-8}$ & $1.4981\times10^{12}$ & $1.237$ & $0.971$\\
	& $0.2662$ & $9.550$ & $0.0290$ & $2.1342$ & $1.3276\times10^{-8}$ & $5.4628\times10^{12}$ & $1.231$ & $0.991$\\[3pt]
ZPG TBL:Exp5	& $0.2690$ & $6.025$ & $0.0343$ & $0.9164$ & $4.5640\times10^{-6}$ & $1.3020\times10^{8}$ & $1.433$ & $0.995$\\
	& $0.4030$ & $9.838$ & $0.0418$ & $2.9777$ & $2.1043\times10^{-6}$ & $5.1505\times10^{8}$ & $1.385$ & $0.999$\\
	& $0.7270$ & $18.950$ & $0.0364$ & $9.4896$ & $1.1911\times10^{-6}$ & $1.4275\times10^{9}$ & $1.364$ & $1.024$\\
	& $0.5120$ & $14.330$ & $0.0347$ & $5.3547$ & $3.9018\times10^{-7}$ & $1.1143\times10^{10}$ & $1.311$ & $0.976$\\
	& $0.5730$ & $17.164$ & $0.0418$ & $9.8128$ & $1.1878\times10^{-7}$ & $9.9160\times10^{10}$ & $1.274$ & $0.947$\\[3pt]
ZPG TBL:Exp6	& $0.6735$ & $17.129$ & $0.0258$ & $5.3334$ & $1.9064\times10^{-6}$ & $6.0272\times10^{8}$ & $1.395$ & $1.059$\\
	& $0.6575$ & $16.676$ & $0.0258$ & $5.0580$ & $1.9624\times10^{-6}$ & $5.7277\times10^{8}$ & $1.397$ & $1.061$\\
	& $0.4946$ & $12.047$ & $0.0258$ & $2.6506$ & $2.8168\times10^{-6}$ & $3.0034\times10^{8}$ & $1.409$ & $1.040$\\
	& $0.7718$ & $19.944$ & $0.0274$ & $7.7779$ & $1.4979\times10^{-6}$ & $9.3461\times10^{8}$ & $1.377$ & $1.040$\\
	& $0.9479$ & $24.991$ & $0.0273$ & $12.2712$ & $1.1661\times10^{-6}$ & $1.4707\times10^{9}$ & $1.364$ & $1.035$\\
	& $0.9978$ & $26.414$ & $0.0274$ & $13.7507$ & $1.0954\times10^{-6}$ & $1.6505\times10^{9}$ & $1.361$ & $1.032$\\
	& $1.1212$ & $29.922$ & $0.0271$ & $17.5895$ & $9.6226\times10^{-7}$ & $2.0951\times10^{9}$ & $1.354$ & $1.025$\\
	& $0.4767$ & $11.528$ & $0.0250$ & $2.3581$ & $3.0515\times10^{-6}$ & $2.5889\times10^{8}$ & $1.410$ & $1.034$\\
	& $0.4961$ & $12.074$ & $0.0252	$ & $2.6051$ & $2.8747\times10^{-6}$ & $2.8803\times10^{8}$ & $1.407$ & $1.035$\\
	& $0.6618$ & $16.777$ & $0.0254	$ & $5.0423$ & $1.9813\times10^{-6}$ & $5.6125\times10^{8}$ & $1.394$ & $1.054$\\
	& $0.6837$ & $17.413$ & $0.0255	$ & $5.4661$ & $1.8882\times10^{-6}$ & $6.1063\times10^{8}$ & $1.392$ & $1.055$\\
	& $0.7761$ & $20.062$ & $0.0248	$ & $7.0926$ & $1.6520\times10^{-6}$ & $7.7242\times10^{8}$ & $1.385$ & $1.058$\\
	& $1.0565$ & $28.084$ & $0.0238	$ & $13.3735$ & $1.1926\times10^{-6}$ & $1.3970\times10^{9}$ & $1.374$ & $1.064$\\
	& $1.1228$ & $29.967$ & $0.0239	$ & $15.3402$ & $1.1049\times10^{-6}$ & $1.6119\times10^{9}$ & $1.370$ & $1.060$\\[3pt]
ZPG TBL:DNS1	& $0.0462^{*}$ & $1.001$ & $2.7714$ & $1.9994$ & $6.6087\times10^{-6}$ & $6.7406\times10^{7}$ & $1.437$ & $0.970$\\
	& $0.0438^{*}$ & $1.000$ & $3.7976$ & $2.7017$ & $4.6366\times10^{-6}$ & $1.2461\times10^{8}$ & $1.424$ & $0.999$\\
	& $0.0422^{*}$ & $0.999$ & $4.6719$ & $3.3136$ & $3.6424\times10^{-6}$ & $1.8955\times10^{8}$ & $1.418$ & $1.026$\\
	& $0.0390$ & $1.002$ & $9.5137$ & $6.9184$ & $1.6122\times10^{-6}$ & $8.1064\times10^{8}$ & $1.381$ & $1.042$\\
	& $0.0384$ & $1.003$ & $10.6300	$ & $7.7192$ & $1.4227\times10^{-6}$ & $1.0101\times10^{9}$ & $1.379$ & $1.055$\\
	& $0.0379$ & $1.003$ & $11.7809	$ & $8.5434$ & $1.2687\times10^{-6}$ & $1.2384\times10^{9}$ & $1.376$ & $1.064$\\
	& $0.0374$ & $1.001$ & $12.9939	$ & $9.3818$ & $1.1401\times10^{-6}$ & $1.4994\times10^{9}$ & $1.373$ & $1.070$\\
	& $0.0371$ & $1.001$ & $14.1632	$ & $10.2361$ & $1.0366\times10^{-6}$ & $1.7824\times10^{9}$ & $1.370$ & $1.071$\\
	& $0.0368$ & $1.000$ & $15.3795	$ & $11.1037$ & $9.4786\times10^{-7}$ & $2.0989\times10^{9}$ & $1.366$ & $1.071$\\[3pt]
ZPG TBL:DNS2	& $0.0385$ & $1.000$ & $0.4949$ & $0.3509$ & $1.6477\times10^{-6}$ & $7.7169\times10^{8}$ & $1.392$ & $1.074$\\
	& $0.0386$ & $1.000$ & $0.4831$ & $0.3424$ & $1.6928\times10^{-6}$ & $7.3532\times10^{8}$ & $1.393$ & $1.073$\\
	& $0.0391$ & $1.000$ & $0.4395$ & $0.3109$ & $1.8856\times10^{-6}$ & $6.0732\times10^{8}$ & $1.397$ & $1.069$
  \end{tabular}
  \caption{Experimental and DNS data of ZPG TBLs from the literature. See table \ref{table:refslinks} for the references and data sources.}
  \label{table:ZPG}
  \end{center}
\end{table}

\begin{table}
  \begin{center}
\def~{\hphantom{0}}
  \begin{tabular}{llll}%{lllllllll}%
      \myalign{c}{Data set} & \myalign{c}{Reference} & \myalign{c}{Digitized, Provided or Data Hyperlink}\\
      \myalign{c}{Code}  & \myalign{c}{paper} & \\[6pt]
Channel:Exp & \cite{schultz2013} & Provided by the authors upon request\\
Channel:DNS1 & \cite{bernardini2014} & \url{http://newton.dma.uniroma1.it/channel/stat/}\\
Channel:DNS2 & \cite{lee2015} & \url{https://turbulence.oden.utexas.edu/channel2015/data/}\\
Pipe:Exp & \cite{zagarola1998JFM} & \url{https://smits.princeton.edu/zagarola/}\\
(with $\dagger$ in table\ref{table:channelpipe}) & &\\
Pipe:Exp & \cite{mckeon2004JFMa} & \url{https://smits.princeton.edu/mckeon/}\\
(without $\dagger$ in table\ref{table:channelpipe}) & &\\
Pipe:DNS1 & \cite{elkhoury2013} & \url{https://kth.app.box.com/v/straightpipestat/}\\
Pipe:DNS2 & \cite{chin2014IJHFF} & \url{https://www.adelaide.edu.au/directory/rey.chin/}\\
ZPG TBL:Exp1 & \cite{harun2013} & Provided by the authors upon request\\       
ZPG TBL:Exp2 & \cite{marusic2015} & Provided by the authors upon request\\
ZPG TBL:Exp3 & \cite{talluru2014} & Provided by the authors upon request\\             
ZPG TBL:Exp4 & \cite{vallikivi2015} & Digitized from figure 3(\emph{b}) of the reference\\      
ZPG TBL:Exp5 & \cite{degraaff2000} & Digitized from figure 2 of the reference\\      
ZPG TBL:Exp6 & \cite{orlu2013} & \url{https://kth.app.box.com/v/TBL-EXP-ZPGTBL/}\\             
ZPG TBL:DNS1 & \cite{jimenez2010} & \url{https://torroja.dmt.upm.es/turbdata/blayers/low\_re/}\\
(with $*$ in table\ref{table:channelpipe}) & & \\      
ZPG TBL:DNS1 & \cite{sillero2013} & \url{https://torroja.dmt.upm.es/turbdata/blayers/high\_re/}\\
(without $*$ in table\ref{table:channelpipe}) & & \\
ZPG TBL:DNS2 & \cite{schlatter2010} & \url{https://www.mech.kth.se/\~pschlatt/DATA/}\\      
        \end{tabular}
  \caption{References and sources for the data listed in tables \ref{table:channelpipe} and \ref{table:ZPG}.}
  \label{table:refslinks}
  \end{center}
\end{table}

\end{document}